\documentclass[useAMS,usenatbib,fleqn]{mnras}

\usepackage{amsmath}
\usepackage{amssymb}
\usepackage[utf8]{inputenc}
\usepackage{xspace}
\usepackage[american]{babel}
\usepackage{times}
\usepackage{graphicx}
\usepackage{color}
\usepackage{multirow}
\usepackage{float}
\usepackage{url}

\newcommand{\ie}{i.\,e.\xspace}

\newcommand{\eg}{e.\,g.\xspace}

\newcommand{\GRBs}{Gamma-Ray Bursts\xspace}

\newcommand{\tv}{\theta_\mathrm{v}}

\newcommand{\tj}{\theta_\mathrm{jet}}

\newcommand{\Ep}{E_\mathrm{peak}}

\newcommand{\toff}{t_{\mathrm{off}}}

\newcommand{\simpropto}{\mathrel{\vcenter{
  \offinterlineskip\halign{\hfil$##$\cr
    \propto\cr\noalign{\kern2pt}\sim\cr\noalign{\kern-2pt}}}}}

\title[Single pulses from off-axis GRBs]{Light curves and spectra from off-axis gamma-ray bursts} 
\author[O. S. Salafia, G. Ghisellini, A. Pescalli, G. Ghirlanda, F. Nappo]{O. S. Salafia$^{1,3}$\thanks{E-mail:
omsharan.salafia@brera.inaf.it (OA Brera Merate), o.salafia@campus.unimib.it (Univ. Milano-Bicocca)}, G. Ghisellini$^{3}$, A. Pescalli$^{2,3}$, G. Ghirlanda$^{3}$, F. Nappo$^{2,3}$\\
$^{1}$Universit\`a degli Studi di Milano-Bicocca, Piazza della Scienza 3, I-20126 Milano, Italy\\
$^{2}$Universit\`a degli Studi dell'Insubria, Via Valleggio, 11, I-22100 Como, Italy\\
$^{3}$INAF - Osservatorio Astronomico di Brera Merate, via E. Bianchi 46, I–23807 Merate, Italy\\
}

\begin{document}
 
 \date{Draft version, \today}
 
 \pagerange{\pageref{firstpage}--\pageref{lastpage}} \pubyear{2015}
 
 \maketitle
 
 \label{firstpage}

 
\begin{abstract}
If gamma-ray burst prompt emission originates at a typical radius, and if material producing the emission moves at relativistic speed, then  the variability of the resulting light curve depends on the viewing angle. This is due to the fact that the pulse evolution time scale is Doppler contracted, while the pulse separation is not. For off-axis viewing angles $\theta_{\rm view} \gtrsim \theta_{\rm jet} + \Gamma^{-1}$, the pulse broadening significantly smears out the light curve variability. This is largely independent of geometry and emission processes.
To explore a specific case, we set up a simple model of a single pulse under the assumption that the pulse rise and decay are dominated by the shell curvature effect. We show that such a pulse observed off-axis is (i) broader, (ii) softer and (iii) displays a different hardness-intensity correlation with respect to the same pulse seen on-axis. For each of these effects, we provide an intuitive physical explanation. We then show how a synthetic light curve made by a superposition of pulses changes with increasing viewing angle. We find that a highly variable light curve, (as seen on-axis) becomes smooth and apparently single-pulsed (when seen off-axis) because of pulse overlap. To test the relevance of this fact, we estimate the fraction of off-axis gamma-ray bursts detectable by \textit{Swift} as a function of redshift, finding that a sizable fraction (between 10\% and 80\%) of nearby ($z<0.1$) bursts are observed with $\theta_{\rm view} \gtrsim \theta_{\rm jet} + \Gamma^{-1}$.
Based on these results, we argue that low luminosity \GRBs are consistent with being ordinary bursts seen off-axis.
\end{abstract}


\section{Introduction}

Despite more than 40 years of observation and modelling, many features of \GRBs (GRBs hereafter) still lack firm and unanimous explanations. 
The diversity and complexity of GRB prompt emission light curves is often used to illustrate the difficulty in the classification of these sources and in the unification of their properties. A natural approach to get insight into such complexity is to look for global and average properties, like flux time integral (\ie \textit{fluence}), total duration, average spectrum, peak flux. Alternatively, one can try to break down the light curve into simpler parts following some pattern. If a fundamental building block was identified, the analysis of single blocks could be the key to the unification and disentanglement of properties of the underlying processes. Many authors \citep[\eg][]{Imhof1974,Golenetskii1983Nature,Norris1986,Link1993,Ford1995,Kargatis1995,Liang1996,Preece1998,Ramirez-Ruiz1999,Lee2000,Ghirlanda2002,2011ApJ...740..104H,Lu2012,Basak2014} performed careful analyses of light curves and time resolved spectra looking for patterns and for hints about such fundamental building blocks. As early as 1983, \citeauthor{Golenetskii1983Nature} found evidence of a correlation between spectral peak energy and photon flux during the decay of pulses. Such correlation was later confirmed by \citet{Kargatis1994ApJ}, \citet{Kargatis1995} and \citet{Borgonovo2001ApJ} and became known as the \textit{hardness-intensity correlation} \citep{Ryde1998}.  \citet{Norris1986} was presumably the first to systematically decompose the light curves into pulses and to look for patterns in the properties of these putative building blocks.
Some years later, \cite{1999ApJ...523..187W} developed tools to calculate the emission from a relativistically expanding jet, including the case of an off-axis viewing angle. \cite{Ioka2001} took advantage of this formulation to model the single pulse, finding that the spectral lag-luminosity and variability-luminosity correlations found by \cite{Norris2000} and \cite{Reichart2001} can be explained as viewing angle effects. The pulse model at that stage assumed emission from a unique radius and from an infinitesimally short time interval (\ie a delta function in radius and time). In the following years, other authors proposed increasingly refined models of the pulse \citep[\eg][]{Dermer2004,Genet2009}, but neglected the possibility for the jet to be observed off-axis.

The viewing angle, \ie the angle between the jet axis and the line of sight, is usually assumed to be smaller than the jet semi-aperture, in which case the jet is said to be \textit{on-axis}. For larger viewing angles, \ie for \textit{off-axis} jets, the flux is severely suppressed because of relativistic beaming. Nevertheless, it can be still above detection threshold if the viewing angle is not much larger than the jet semi-aperture, especially if the burst is at low redshift. In \citet{2015MNRAS.447.1911P} we have shown that off-axis jets might indeed dominate the low luminosity end of the observed population.

The idea that nearby low luminosity GRBs could be off-axis events has been a subject of debate since the observation of GRB980425. \citet{Soderberg2004} rejected such possibility, based on radio observations of GRB980425 and GRB031203, but soon later \citet{Ramirez-Ruiz2005} presented an off-axis model for the afterglow of GRB031203 which seems to fit better the observations (including radio) with respect to the usual on-axis modelling. Using the same off-axis afterglow model, \citet{Granot2005a} extended the argument to two X-Ray Flashes, thus including them in the category of off-axis GRBs. Based on prompt emission properties, an off-axis jet interpretation of X-Ray Flashes had been already proposed by \cite{Yamazaki2002} and \cite{Yamazaki2003a}, following the work by \cite{Ioka2001}. \citet{2006MNRAS.372.1699G} argued that the off-axis interpretation of GRB031203 and GRB980425 is not practicable, because their true energy would then be on the very high end of the distribution, implying a very low likelihood when combined with the low redshift of these two events. More recently, the idea that such events are members of a separate class \citep[\eg][]{Liang2007,Zhang2008,He2009,2011ApJ...739L..55B,Nakar2015} has gained popularity. Our results about the GRB luminosity function \citep{2015MNRAS.447.1911P}, though, still point towards the unification of these events with ordinary GRBs based on the off-axis viewing angle argument. With the present work we address the issue from another point of view, by focusing on the apparently single pulsed, smooth behaviour of prompt emission light curves of these bursts, trying to figure out if such behaviour is expected in the case of an off-axis viewing angle.

The structure of the paper is as follows: in section \S\ref{sec:pulses-building-blocks} we explain why an off-axis GRB is always less variable than the same GRB seen on-axis; in \S\ref{sec:pulse-light-curves} we discuss the main assumptions of our simple pulse model and we present the predictions for an on-axis (\S\ref{sec:onaxis-jet}) and an off-axis observer (\S\ref{sec:offaxis-jet}). In \S\ref{sec:multipulse} we build a superposition of pulses (to represent a synthetic prompt emission light curve) and show how its properties change with increasing off-axis viewing angle, comparing them with those found in light curve time resolved spectral analysis. As expected, the model predicts that variability is suppressed in off-axis GRBs because of pulse broadening and overlap. In \S\ref{sec:n_offaxis} off-axis GRBs are shown to be a significant fraction of nearby observed GRBs. Based on the obtained results, we conclude (\S\ref{sec:low-lum-grbs}) that light curves of low luminosity GRBs are consistent with the off-axis hypothesis. We then summarize and draw our conclusions in \S\ref{sec:conclusions}.

\section{Pulses: building blocks of GRB light curves}\label{sec:pulses-building-blocks}

\subsection{Pulse overlap and light curve variability}\label{sec:naive}
\begin{figure}
 \includegraphics[width=\columnwidth]{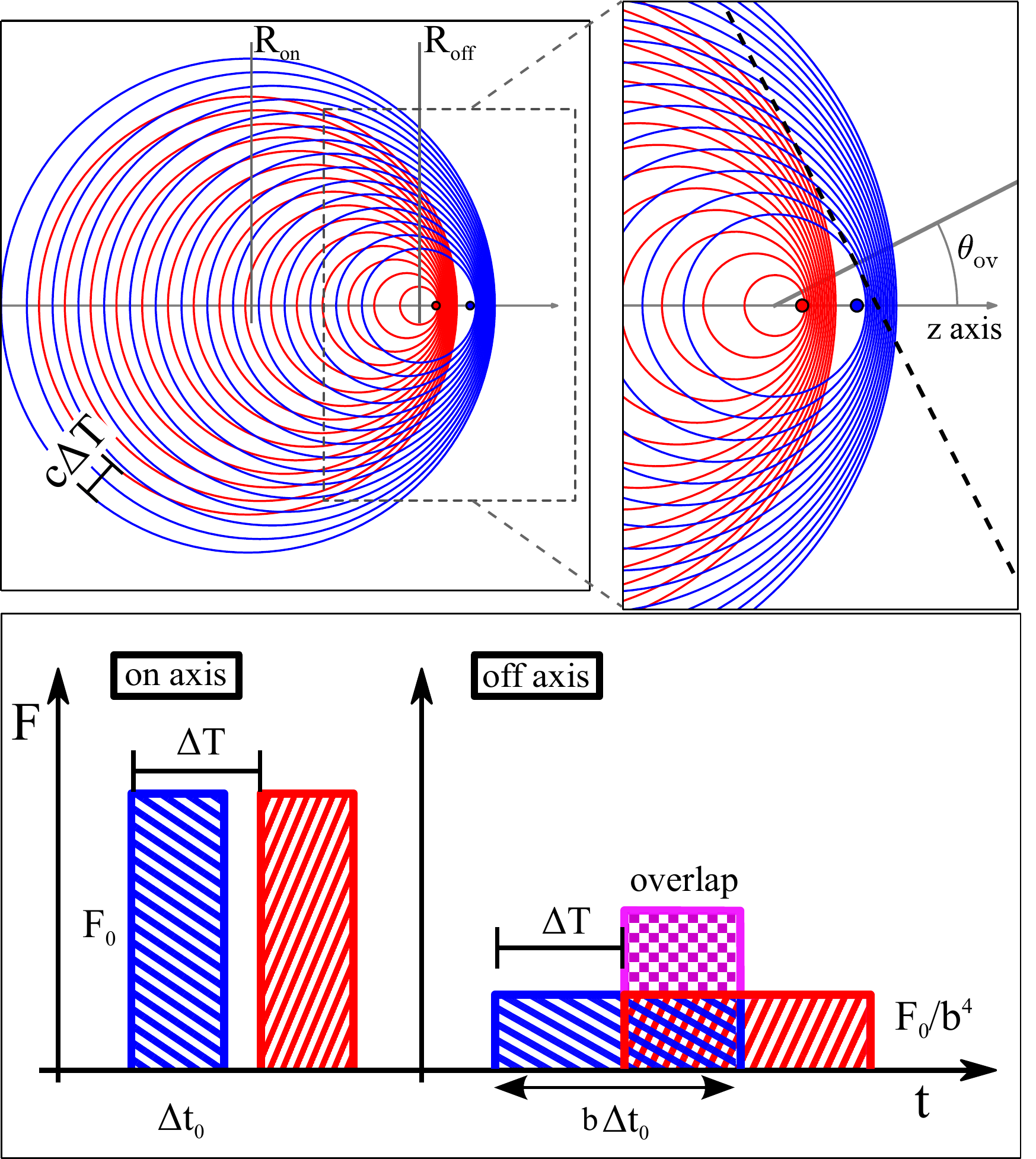}
 \caption{\textbf{Upper left panel}: two point sources (blue and red dots) move at equal constant speed along the $z$ axis, separated by a distance $\beta c \Delta T$. Each starts emitting at $z=R_{\rm on}$ and stops emitting at $z=R_{\rm off}$.  The blue and red circles represent wavefronts of the emitted light. The first blue wavefront and the first red wavefront reach any observer with a time difference $\Delta T$. \textbf{Upper right panel}: close up. Depending on the viewing angle $\theta_{\rm v}$, a distant observer sees the blue and red signal separated ($\theta_{\rm v}<\theta_{\rm ov}$) or overlapped ($\theta_{\rm v}>\theta_{\rm ov}$). The angle $\theta_{\rm ov}$ is the angle between the $z$ axis and the normal to a plane tangent to both the first red wavefront and the last blue wavefront. \textbf{Lower panel}: sketch of the bolometric light curve as seen by on-axis ($\theta_{\rm v} = 0$) and off-axis ($\theta_{\rm v}>\theta_{\rm ov}$) observers. Letting $b=(1-\beta\cos\theta_{\rm v})/(1-\beta)$, the single pulse flux as measured by the off-axis observer is decreased by a factor $b^4$ with respect to the on-axis one, while the duration is increased by a factor $b$. The pulse separation $\Delta T$, though, does not depend on the viewing angle, being the emission time difference at a fixed radius. This causes the pulses to overlap as seen by the off-axis observer. \label{fig:pointsources}}
\end{figure}

In a highly variable light curve, pulses must be short and not overlap too much. If pulses are produced at a typical radius by material moving close to the speed of light, then the amount of overlap can depend on the viewing angle. To see this, consider two point sources moving at equal constant speed $\beta c$ along the $z$ axis, separated by a distance $\beta c \Delta T$, as in Fig.~\ref{fig:pointsources}. Each source starts emitting at radius $R_{\rm on}$ and stops emitting at $R_{\rm off}$. An observer along the $z$ axis (viewing angle $\tv=0$) sees two separated pulses of equal duration $\Delta t_0$ and peak flux $F_0$, the second starting a time $\Delta T$ after the start of the first. Because of relativistic Doppler effect, an observer with another $\theta_{\rm v}\neq 0$ measures a lower (bolometric) peak flux $F = F_0/b^4 $ and a longer pulse duration $\Delta t = b\, \Delta t_0$, where $ b=(1-\beta\cos\theta_{\rm v})/(1-\beta)$ is the ratio of the on-axis relativistic Doppler factor $\delta(0)=\Gamma^{-1}\left(1-\beta\right)^{-1}$ to the off-axis one $\delta(\tv)=\Gamma^{-1}\left(1-\beta\cos\tv\right)^{-1}$ \citep{Rybicki1979,ghisellini_book_2013}. The difference in pulse start times $\Delta T$, on the other hand, is not affected by the viewing angle, because the emission of both pulses begins at the same radius: it can be thought of as emission from a source at rest (for what concerns arrival times). The pulses overlap if $\Delta t > \Delta T$, which corresponds to $\tv > \theta_{\rm ov} \approx \Gamma^{-1}\sqrt{\Delta T/\Delta t_0  -1}$.
Consider the case in which the pulse separation is equal to the pulse duration, \ie $\Delta T = 2\Delta t_0$ and $\theta_{\rm ov} \approx \Gamma^{-1}$. Increasing the viewing angle, the amount of pulse overlap increases, reaching half of the pulse width as soon as $b=4$, which corresponds to $\tv \approx \sqrt{3}\Gamma^{-1}$. With this viewing angle, the flux of the single pulse is reduced by $b^4=256$, but the flux in the overlapped region is higher by a factor of two, so that the peak flux effectively decreases by $128$.

 \begin{figure}
  \includegraphics[width=\columnwidth]{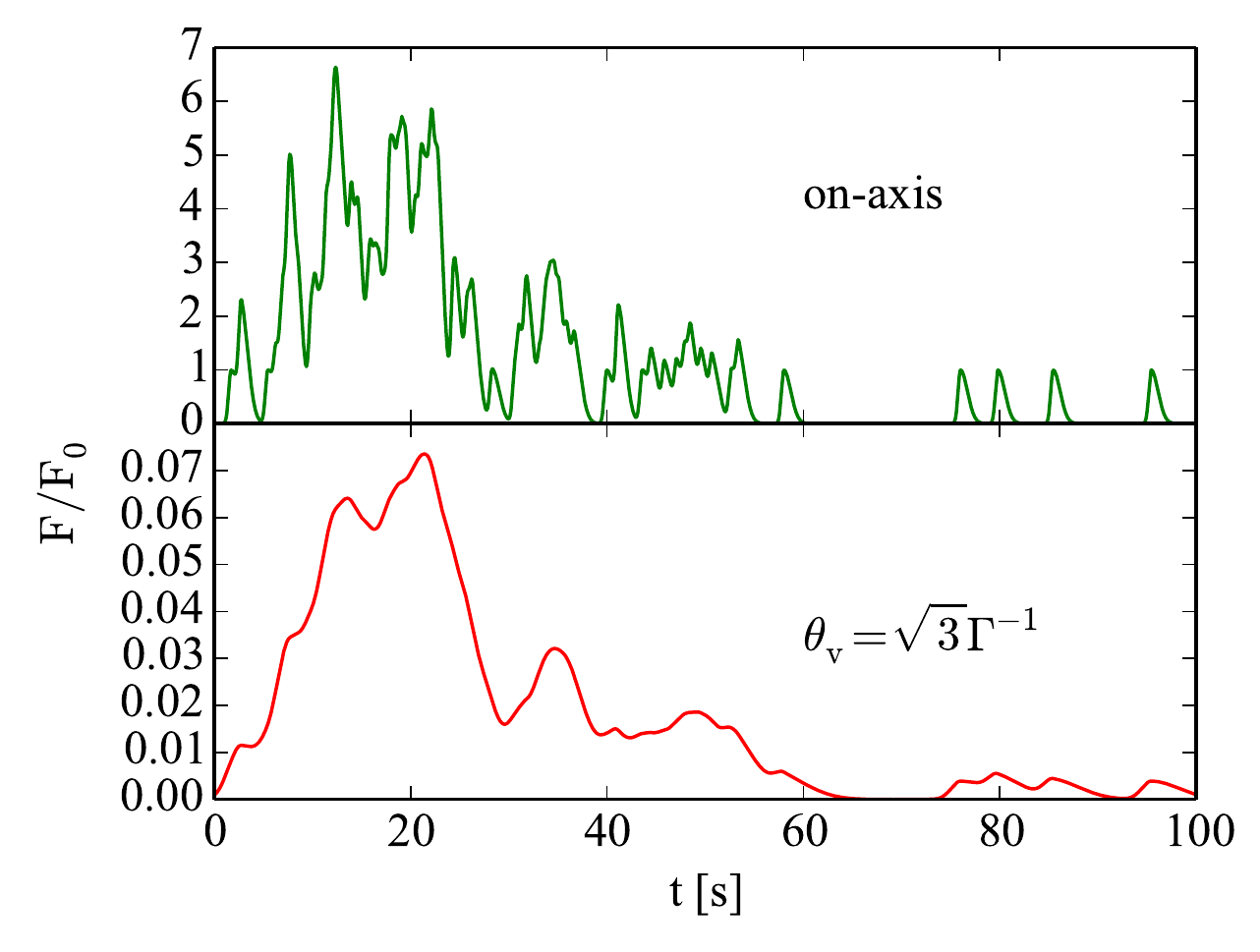}
  \caption{Example light curves constructed by superposition of pulses. All pulses are equal. The pulse shape is a double-sided Gaussian \citep{Norris1996}, which is a common phenomenological description of GRB pulse shapes. The peak flux is $F_0$, and the rise to decay time ratio is 1:3. The start times of the pulses are the same for the two light curves and have been sampled from a log-normal distribution with mean $20\,\rm{s}$ and sigma $0.35\,\rm{dex}$. Pulses in the lower light curve are broadened by a factor of $4$ and their flux is lowered by a factor of $256$ with respect to the upper light curve, which corresponds to the effect of an off-axis viewing angle $\tv = \sqrt{3}\Gamma^{-1}$ as discussed in the text.\label{fig:multipulse-simple}}
 \end{figure}

The purpose of this simple argument is to show that if pulses are produced by material moving at relativistic speed, and if a typical emission radius exists, then the apparent variability of the light curve can be significantly smeared out by pulse overlap as seen by an off-axis observer (see also Fig.~\ref{fig:multipulse-simple}). The viewing angle needed for this to happen is still small enough for the flux not to be heavily suppressed by relativistic (de-)beaming. One may argue that the probability to have a viewing angle in the right range for this to happen without falling below the limiting flux of the instrument is vanishingly small. To address this point, in \S\ref{sec:n_offaxis} we give an estimate of the rate of such events, showing that a significant fraction ($\sim 40\%$) of nearby bursts ($z<0.1$) are likely observed with $\tv>\tj+\Gamma^{-1}$.

Being based solely on geometry and relativity, the above argument does not rely on a specific scenario, \eg internal shocks. Any model in which photons are produced at a typical radius, being the photospheric radius \citep[\eg subphotospheric dissipation models like those described in][]{2005ApJ...628..847R,Giannios2006,Beloborodov2010} or beyond \citep[\eg magnetic reconnection models,][]{Lazarian2003,Zhang2011} eventually must take into account the pulse overlap as seen by off-axis observers.


\subsection{Pulses in the internal shock scenario}
The pulse width in GRB light curves is roughly constant throughout the burst duration \citep{Ramirez-Ruiz1999}. The internal shock scenario \citep{Rees1994} provides a natural framework for the understanding of this kind of behaviour. In this scenario, discontinuous activity in the central engine  produces a sequence of shells with different Lorentz factors. When faster shells catch up with slower ones, shocks develop and particles are heated. If the plasma is optically thin and some magnetic field is present, the energy gained by the electrons is promptly and efficiently radiated away by synchrotron (and inverse Compton) emission. Each pulse is thus the result of the merger of two shells beyond the photospheric radius $R_{ph}$. The strength of the shock, and thus the efficiency of the electron heating, depends strongly on the relative Lorentz factor of the merging shells (a radiative efficiency of a few percent is achieved only for $\Gamma_{\rm rel}\gtrsim 3$, \citeauthor{Lazzati1999} \citeyear{Lazzati1999}). Shell pairs with small relative Lorentz factors merge later (they need more time to catch up with each other), thus the highest efficiency is achieved for shells merging just after the photospheric radius. This explains, within this framework, why the typical pulse width is not seen to grow with time: the bulk of the emission happens at a fixed radius, regardless of the expansion of the jet head.

\subsection{Time scales}
\label{sec:timescales}
Three main time scales arise in the internal shock scenario:
\begin{itemize}
 \item the electron cooling time $\tau_{cool}$, \ie the time needed by electrons to radiate away most of the energy gained from the shock;
 \item the angular time scale $\tau_{ang}$, \ie the difference in arrival time between photons emitted at different latitudes;
 \item the shell crossing time $\tau_{sc}$, \ie the time needed for the two shells to merge.

\end{itemize}

The electron cooling time scale, as measured in the lab frame, is $\tau_{\rm cool} \sim \Gamma^{-1} \gamma/\dot{\gamma}$, where $\Gamma$ is the bulk Lorentz factor, $\gamma$ is the typical electron Lorentz factor as measured in the comoving frame, and $\dot{\gamma}$ is the cooling rate. For synchrotron emission, it is of the order of $\tau_{\rm{cool}}  \sim 10^{-7}$~s for typical parameters\footnote{by typical parameters we mean $\Gamma = 100$, $\Gamma_{\rm rel} = \text{a few}$, $U_{\rm rad}=U_B$, a typical synchrotron frequency of $1$ MeV and we assume equipartition. See \citet{Ghisellini2000} and references therein for a complete treatment.} \citep{Ghisellini2000}.

The angular time scale arises when one takes into account the arrival time difference of photons emitted at the same time by parts of the shell at different latitudes. It is defined as the arrival time difference between a pair of photons, one emitted at zero latitude and the other at $\Gamma^{-1}$ latitude. Given a typical photospheric radius \citep{daigne-thermal-precursors-2002} $R_{ph}\sim 10^{12}$~cm, this difference is $\tau_{ang} \sim R/\Gamma^2 c \approx 3 \times 10^{-3}$~s~$R_{12}/\Gamma_2^{2}$ (we adopt the notation $Q_x = Q/10^x$ in cgs units). 

The shell crossing time is $\tau_{sc}\sim w/c$, where $w$ is the typical shell width. Being linked to the central engine activity, one may assume $w$ to be of the order of a few Schwarzschild radii. The Schwarzschild radius of a $5\,M_{\odot}$ black hole is $R_{s}\approx 1.5\times 10^{6}$~cm, thus an estimate might be $\tau_{sc}\sim 5\times 10^{-5}$~s~$w_6$. In this case, we have $\tau_{\rm ang}>\tau_{\rm sc}$, \ie the effect of shell curvature dominates over (\ie smears out) intrinsic luminosity variations due to shock dynamics, which take place over the $\tau_{\rm sc}$ time scale or less.

Temporal analysis of GRB light curves, though, along with simple modelling of internal shocks \citep{Nakar2002a,Nakar2002b}, seem to indicate that the shell width must be comparable to the initial shell separation. Taking the two as equal, the time needed for two shells to collide is the same as the shell crossing time, and thus the shell merger is completed within a doubling of the radius. In this case, the shell crossing time and the angular time scale are the same \citep{piran-physics_of_grbs04}. This means that details of the pulse shape and spectral evolution cannot be explained as just being due to the shell curvature effect. Indeed, discrepancies between predictions based on shell curvature only and observations have been pointed out \citep[\eg][]{Dermer2004}.

Nevertheless, the description of the pulse in terms of shell curvature qualitatively reproduces the main features of many long GRB pulses, namely the fast rise and slower decay, the hard-to-soft spectral evolution, and the presence of a hardness-intensity correlation \citep{Ryde2002}. For this reason, since we focus on the effect of the viewing angle rather than on details of the pulse, in what follows we set up a simple model of the pulse based on the shell curvature effect only.

\section{Pulse light curves and time dependent spectra}\label{sec:pulse-light-curves}

\subsection{Main assumptions}\label{sec:assumptions}
Based on the arguments outlined in \S\ref{sec:timescales}, we assume that the variation of the flux seen by the observer during a single pulse is due only to the angular time delay described above. The luminosity $L$ of the shell is assumed constant during an emission time $T$ and zero before and after this time interval. The emitting region is assumed geometrically and optically thin. The emitted spectrum, as measured by a locally comoving observer, is assumed to be the same for any shell fluid element.

\cite{1999ApJ...523..187W} and other authors already provided the necessary formulas for the computation of the pulse shape in this case. For the ease of the reader, and for the purpose of developing an intuitive physical description of the results, though, we will go through some details of the derivation anyway, hereafter and in the appendix.

Let the radius of the shell be $R$ at the beginning of the emission and $R+\Delta R$ at the end of it. The bolometric flux $F(t)$ (specific flux $F_\nu(t)$) is computed by integration of the intensity $I$ (specific intensity $I_\nu$) over the appropriate equal arrival time surface $S(t)$, namely
\begin{equation}
 F(t) = \int_{S(t)}I(s)\,ds/r^2
 \label{eq:flux-moving-source-text}
\end{equation}
where $r$ is the distance between the element $ds$ of $S(t)$ and the observer. 

Assuming isotropic emission in the comoving frame, in the approximation of infinitesimal shell thickness, the intensity is related to the comoving one (primed quantities throughout the paper will always refer to the comoving frame) through $I = \delta^4 I'$ (or $I_\nu(\nu) = \delta^3 I_\nu'(\nu/\delta)$ for the specific intensity). Note that the constant luminosity assumption implies $I'\propto R^{-2}$: this is consistent if the number of emitting particles is constant despite the increase of the surface area with the expansion. It would not be appropriate \eg for external shocks, where the number of emitting particles instead increases with increasing surface area.

\subsection{Equal arrival time surfaces}

\subsubsection{A sphere}\label{sec:eats-sphere}

\begin{figure}
 \begin{center}
 \includegraphics[width=\columnwidth]{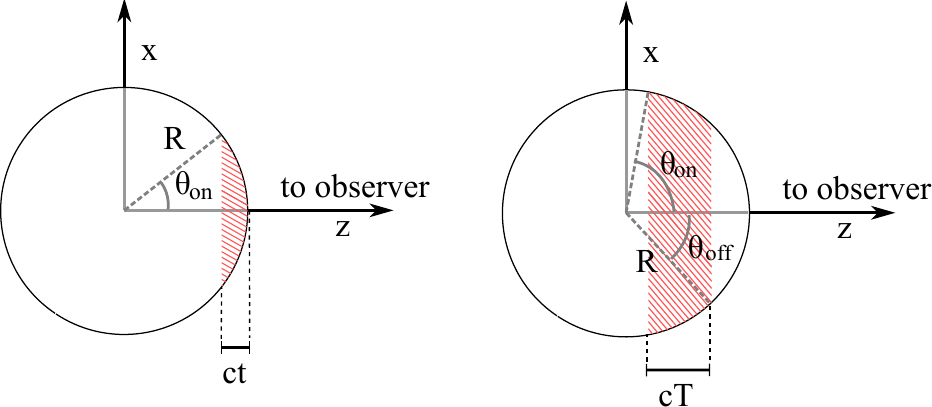} 
 \end{center}
 \caption{\label{fig:geom1} A sphere starts emitting radiation at $t=t_0$ and stops at $t=t_0+T$. The line of sight of a distant observer is parallel to the z axis. \textbf{Left:} a time $t < T$ after the arrival of the first photon, the observer has received radiation from the portion of the sphere with $z>R-c t=R\cos\theta_{\rm on}$; \textbf{Right:} later when $t > T$, the observer has stopped receiving radiation from the portion of the sphere with $z>R-c(t-T)=R\cos\theta_{\rm off}$. Thus the effective emitting surface is the portion of the sphere with $R\cos\theta_{\rm on} < z < R\cos\theta_{\rm off}$. }
\end{figure}

Consider a sphere of radius $R$. The surface of the sphere starts emitting electromagnetic radiation at $t=t_0$ and stops suddenly at $t=t_0+T$ (as measured in the inertial frame at rest with respect to the centre of the sphere). Emitted photons reach a distant observer at different arrival times. Let the line of sight be parallel to the $z$ axis (as in Fig.~\ref{fig:geom1}). The first photon to reach the observer is the one emitted at $t=t_0$ from the tip of the sphere at $z = R$. Let $t = 0$ be its arrival time as measured by the observer. A photon emitted at the same time $t=t_0$ by a point of the surface at $z = R\cos\theta_{\rm on}$ reaches the observer at a later time $t = R(1-\cos\theta_{\rm on})/c$. Thus, despite the surface turned on all at the same time $t=t_0$, at a given time $t$ the observer has received radiation only from the portion with $z/R > \cos\theta_{\rm on} = 1-c t/R$ (left panel of Fig.~\ref{fig:geom1}). This can be visualized as each point on the sphere being turned on by the passage of a plane traveling in the $-z$ direction with speed $c$, starting from $z = R$ at $t=0$. The same reasoning applies to the turning-off of the sphere: each point is turned off by the passage of a plane traveling in the $-z$ direction with speed $c$, starting from $z=R$ at $t=T$. As a result, if $T<R/c$, at some time $t$ the observer will ``see'' only the portion of sphere comprised between $\cos\theta_{\rm on} = 1-c t/R$ and $\cos\theta_{\rm off}=1-c (t-T)/R$ (right part of Fig.~\ref{fig:geom1}). Thus, the EATS at time $t$ is this portion of the sphere.

\subsubsection{An expanding sphere}\label{sec:eats-exp-sphere}

\begin{figure}
 \includegraphics{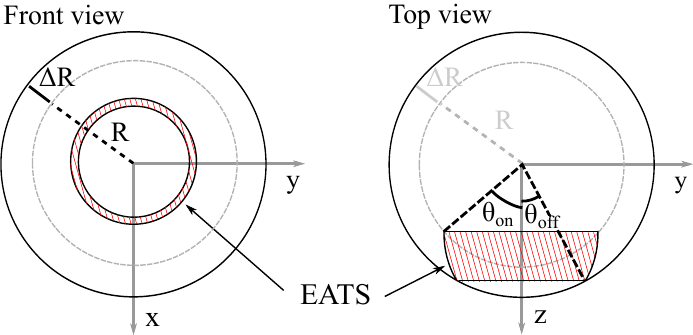}
 \caption{\label{fig:geometry-eats} The shaded regions represent the equal arrival time surface (EATS) of the expanding sphere at $t>t_{\rm off}$. The line of sight is parallel to the $z$-axis. The sphere started emitting when its radius was $R$, and stopped when it was $R+\Delta R$. }
\end{figure}

If the sphere is expanding, the above argument is still valid, with some modification. The radius now is $R(t) = R + \beta c (t-t_0)$ so that the lighting up takes place at $R(0)=R$ and the turning-off at $R(T)=R+\beta c T\equiv R + \Delta R$. Since the lighting up happens all at the same radius, the angle $\theta_{\rm on}$ up to which the observer sees the surface on is still given by $\cos\theta_{\rm on} = 1 - c t/R$ (the photons emitted at $t=t_0$ all come from the sphere with radius $R$). Since the sphere is expanding, its surface ``runs after'' the emitted photons, causing the arrival time difference between the first and the last photon to contract. In particular, for the first and last photon emitted from $z=R(t)$, the arrival time difference is $\toff = T(1-\beta)=T/(1+\beta)\Gamma^2$ where $\Gamma=(1-\beta^2)^{-1/2}$ is the Lorentz factor of the expansion. For this reason, the angle up to which the observer sees the surface turned off is given by $\cos\theta_{\rm off} = 1 - c(t-\toff)/(R + \Delta R)$.

The resulting geometry is not spherical (see Fig.~\ref{fig:geometry-eats}), but the assumption of constant luminosity greatly simplifies the mathematical treatment in that it allows one to perform all integrations over angular coordinates only.

\subsection{On-axis jet}\label{sec:onaxis-jet}

\begin{figure}
 \includegraphics[width=\columnwidth]{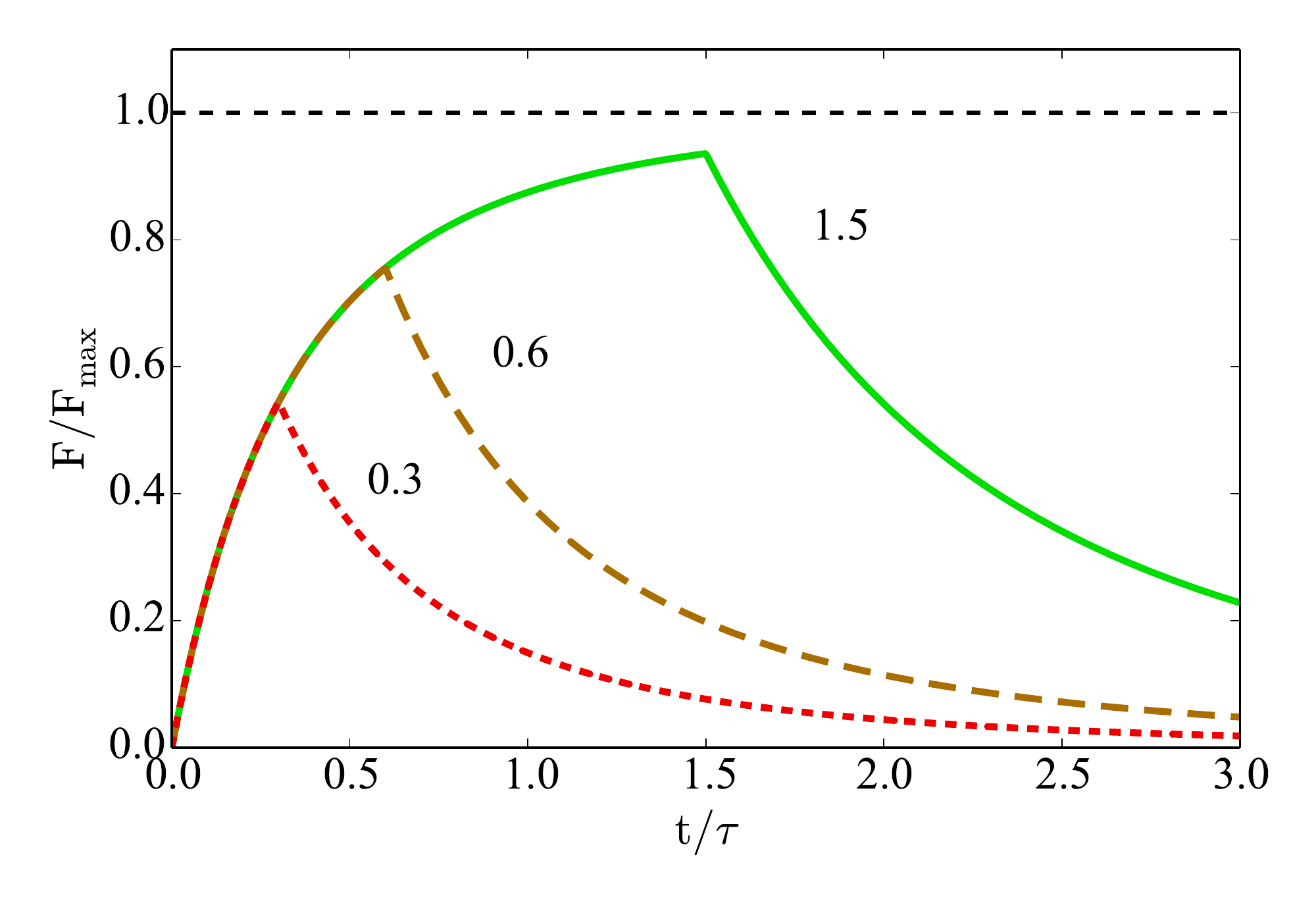}
 \caption{\label{fig:sphere-pulse} Bolometric light curves of three pulses from an expanding sphere. The flux is normalized to $F_{\rm max}$ and the observer time is in units of $\tau$ (see the text for the definition of these quantities). The ratio of $\Delta R$ to $R$ is given near each curve. The black dashed line represents the saturation flux, which is reached if $T\gg R/c$, or equivalently if $\Delta R \gg R$.}
\end{figure}

\begin{figure}
 \includegraphics[width=\columnwidth]{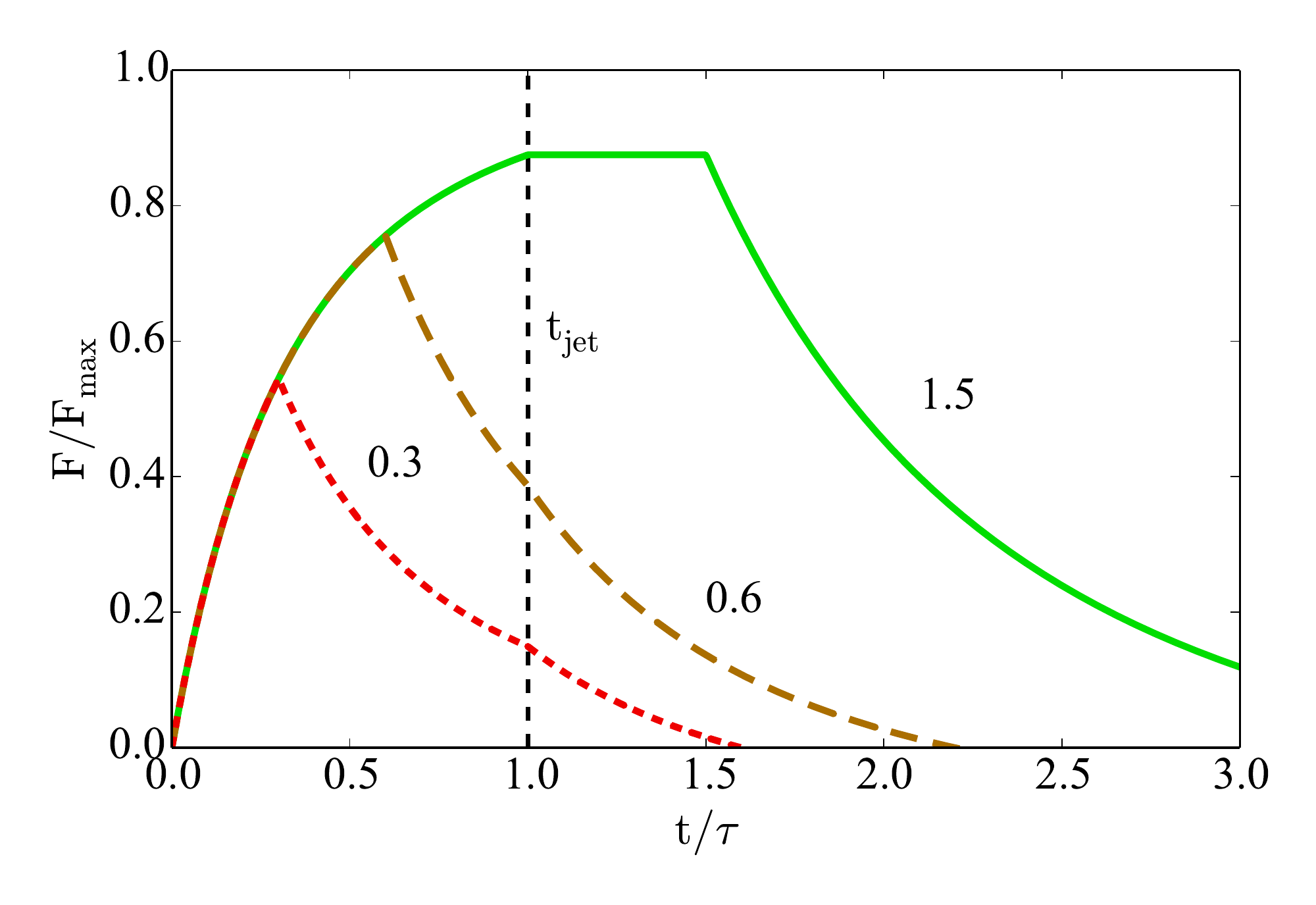}
 \caption{\label{fig:pulse-thj} Bolometric light curves of three pulses from an on-axis jet with $\tj = 1/\Gamma$. The ratio of $\Delta R$ to $R$  is reported near each curve. The black dashed line represents the time $t_{\rm jet}$ at which the jet border first comes into sight. In this case $t_{\rm jet}$ equals $\tau$.}
\end{figure}

A radially expanding (homogeneous) jet seen on-axis is indistinguishable from an expanding sphere as long as its half opening angle $\tj$ is much larger than $1/\Gamma$ \citep[\eg][]{rhoads-balloon97}. As a corollary, the same holds if the viewing angle $\tv$ is small enough so that the angular distance $\tj-\tv$ of the line of sight from the jet border is still much larger than $1/\Gamma$. Since the typical expected Lorentz factor of GRB jets is $\Gamma\sim 100$, this means that a viewing angle a few $0.01$ radians smaller than $\tj$ allows one to consider the jet practically on-axis. On the other hand, if the jet is very narrow, or if the Lorentz factor is low enough (\ie if $\tj$ is comparable with $1/\Gamma$), then the finite half opening angle must come into play. Figure~\ref{fig:sphere-pulse} shows light curves (Eqs. \ref{eq:spherepulse1} \& \ref{eq:spherepulse2}) of pulses from an expanding sphere, or equivalently from a jet with $\tj\gg 1/\Gamma$; Figure~\ref{fig:pulse-thj} instead shows light curves (Eqs.~\ref{eq:pulse-thj-1-appendix}~\&~\ref{eq:pulse-thj-2-appendix}) of pulses from an on-axis jet with $\tj=1/\Gamma$. 

Such light curves can be computed analytically within the assumptions stated in \S\ref{sec:assumptions}. Some natural scales emerge during the derivation: 
\begin{enumerate}
 \item the angular time scale 
 $$
 \tau \equiv \dfrac{R}{\beta c(1+\beta)\Gamma^2}
 $$
 \item the pulse peak time
 $$
 t_{\rm peak} \equiv \dfrac{\Delta R}{\beta c(1+\beta)\Gamma^2}
 $$
 \item the pulse saturation flux 
 $$
 F_{\rm max} \equiv \dfrac{2\pi\,R^2}{ 3 d^2}\dfrac{(1+\beta)^3}{\beta}\Gamma^2\,I_0'
 $$
\end{enumerate}
 where $I_0'$ is the comoving bolometric intensity and $d$ is the distance of the jet from the observer.

\noindent
In the spherical case, before the peak ($t\leq t_{\rm peak}$), the bolometric flux rises as
\begin{equation}
 \frac{F(t)}{F_{\rm max}} =  1 - \left(1+\frac{t}{\tau} \right)^{-3}
 \label{eq:spherepulse1}
 \end{equation}
then ($t > t_{\rm peak}$)  it decreases as
\begin{equation}
   \frac{F(t)}{F_{\rm max}} = \left(1+\frac{t-t_{\rm peak}}{\tau + t_{\rm peak}} \right)^{-3} - \left(1+\frac{t}{\tau} \right)^{-3}
   \label{eq:spherepulse2}
\end{equation}

When the finite jet opening angle $\tj$ is taken into account, the light curve is given instead by Eqs.~\ref{eq:pulse-thj-1-appendix}\&~\ref{eq:pulse-thj-2-appendix}.

\subsubsection{Spectra and hardness-intensity correlation}

\begin{figure}
 \includegraphics{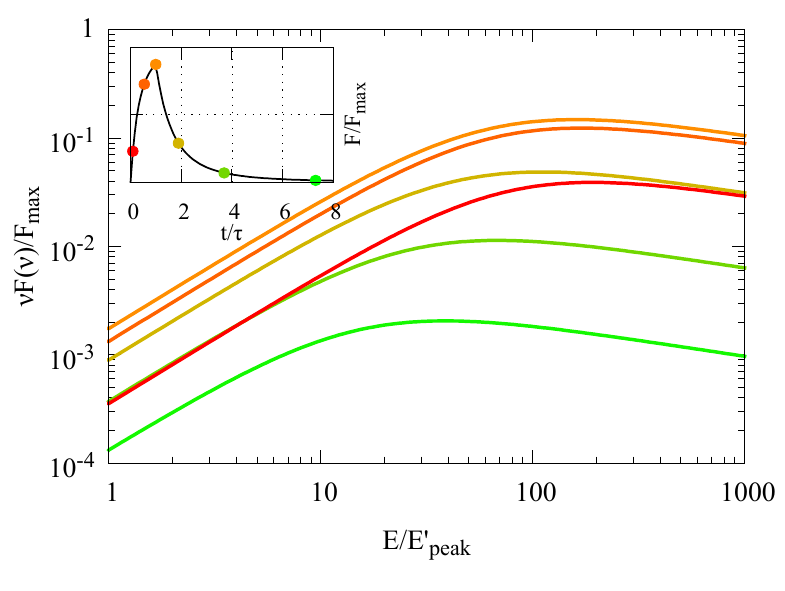}
 \caption{\label{fig:on-axis-spectra} Spectra at different times of a pulse from a jet seen on-axis, with $\Gamma = 100$ and $\Delta R = R$. The comoving spectral shape is given in Eq.~\ref{eq:comov-spec-shape}. The coloured circles in the inset show at which point in the pulse each spectrum (identified by the colour) was calculated. }
\end{figure}

\begin{figure}
 \includegraphics{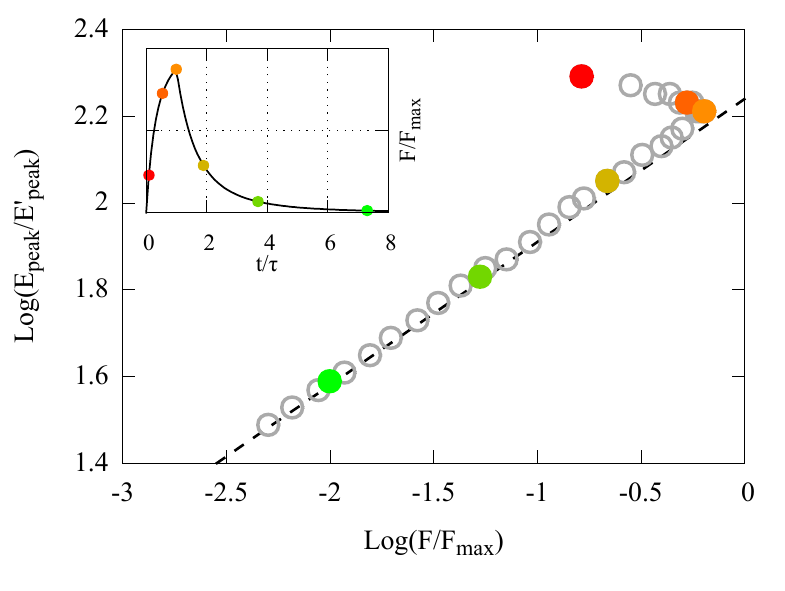}
 \caption{\label{fig:yone-time-res} Peak of the observed spectrum versus the bolometric flux, for a pulse with $\Gamma = 100$ and $\Delta R = R$. A clear hardness-intensity correlation is present. The slope of the black dashed line is $1/3$. The inset is the same as in Fig.~\ref{fig:on-axis-spectra}.}
\end{figure}

Since we are mainly interested in how the peak of the observed spectrum evolves with time, we assume a simple form of the comoving spectral shape, namely
\begin{equation}
 I_\nu'(\nu') = n(a,b)\dfrac{I_0'}{\nu_0'}\left[\left(\dfrac{\nu'}{\nu_0'}\right)^{-a} + \left(\dfrac{\nu'}{\nu_0'}\right)^{-b}\right]^{-1}
 \label{eq:comov-spec-shape}
\end{equation}
where $n(a,b)$ is a normalization constant which depends upon the high and low spectral indices $a$ and $b$; clearly $I_\nu' \propto \nu'^{a}$ for $\nu'\ll \nu_0'$ and $I_\nu' \propto \nu'^{b}$ for $\nu'\gg \nu_0'$. If $a>0$ and $b<-1$, the normalization $n(a,b)$ can be defined so that
\begin{equation}
 I_0' = \int_0^\infty I_\nu'(\nu')\,d\nu'
\end{equation}
The break frequency $\nu_0'$ is related to the comoving $\nu' F_\nu'$ peak energy $E_{\rm peak}'$ through
\begin{equation}
 E_{\rm peak}' = \left(-\dfrac{a+1}{b+1}\right)^\frac{1}{a-b} h\nu_0'
\end{equation}
where $h$ is Planck's constant.
All the examples in this paper will assume the above comoving spectral shape, with $a=0.2$ and $b=-1.3$, which represent average high and low spectral indices of Fermi GRB spectra \citep{2011A&A...530A..21N}.

Figure~\ref{fig:on-axis-spectra} shows spectra from an on-axis jet at six representative times, computed using Eq.~\ref{eq:off-axis-spectrum}. The evolution is clearly hard-to-soft (\ie the peak energy decreases monotonically with time), and the low and high energy spectral indices are the same as those of the comoving spectrum. Figure~\ref{fig:yone-time-res} shows that after the peak of the light curve the peak energy $\Ep$ of the observed $\nu F_\nu$ spectrum varies with the bolometric flux $F$ following roughly $\Ep \propto F^{1/3}$, \ie the model predicts a hardness intensity correlation with index $1/3$ during the decay of the pulse.
Let us interpret these results:
\begin{enumerate}
 \item \textit{Pulse rise:}
the maximum of $\Ep(t)$ is at the very beginning of the pulse, when only a small area pointing directly towards the observer (the ``tip'' of the jet at zero latitude) is visible. As the visible area increases, less beamed contributions from parts at increasing latitude come into sight, reducing $\Ep$ slightly. 

\item \textit{Pulse decay:}
after the pulse peak, the tip of the jet turns off, causing $\Ep$ to drop. At this time the visible part of the jet is an annulus (see Fig.~\ref{fig:geometry-eats}): the spectral peak is determined mainly by the maximum Doppler factor $\delta_{\rm max}(t) = \Gamma^{-1}\left[1-\beta\cos\theta_{\rm off}(t)\right]^{-1}$ of the visible area, which corresponds to the innermost circle of the annulus, so that $\Ep \sim \delta_{\rm max} E_{\rm peak}'$. The flux $F$ in turn decreases approximately as $\delta_{\rm max}^4$ times the angular size of the annulus. The latter is proportional to $\cos\theta_{\rm off}-\cos\theta_{\rm on}$, which can be shown to be
\begin{equation}
 \cos\theta_{\rm off}-\cos\theta_{\rm on} \approx \frac{\Delta R}{R}\frac{1}{\beta \Gamma \delta_{\rm max}}
\end{equation} 
As a result, we have that $F\propto \delta_{\rm max}^3$, which explains why $\Ep \propto F^{1/3}$.

\end{enumerate}

\subsection{Off-axis jet}\label{sec:offaxis-jet}

\begin{figure}
 \begin{center}
 \includegraphics[width=\columnwidth]{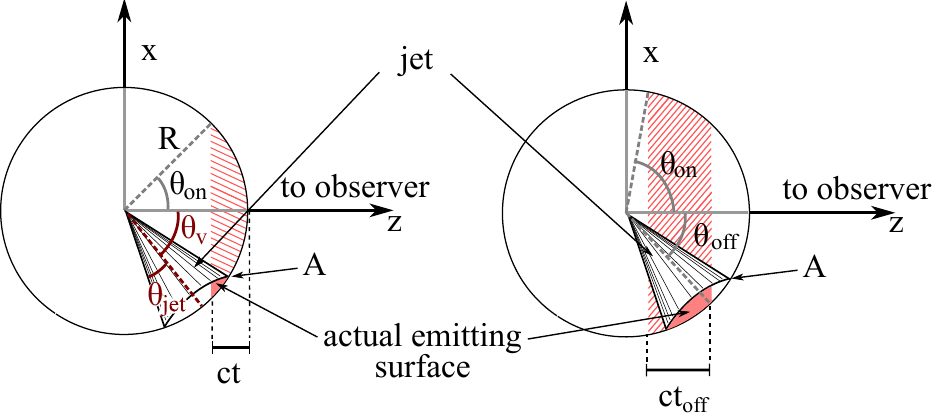} 
 \end{center}
 \caption{\label{fig:geom6}  The off-axis jet can be thought of as being part of an expanding sphere. For simplicity, the EATS of the expanding sphere (hatched area) is represented as in Figure~\ref{fig:geom1}, but it is actually the same as in Figure~\ref{fig:geometry-eats}. The actual EATS of the jet is the interception between the jet surface and the sphere EATS.}
\end{figure}

 Also the off-axis jet ($\tv>\tj$) can be treated using the formalism introduced by \cite{1999ApJ...523..187W}. In the appendix (\S\ref{sec:offaxis-appendix}) we give an alternative derivation based on the idea that the off-axis jet can be thought of as being part of an expanding sphere, and that we can work out the proper EATS as the intersection between the expanding sphere EATS and the jet surface. 
 
 \subsubsection{A longer pulse duration}\label{sec:longer-duration}
 
 \begin{figure}
 \begin{center}
 \includegraphics[width=\columnwidth]{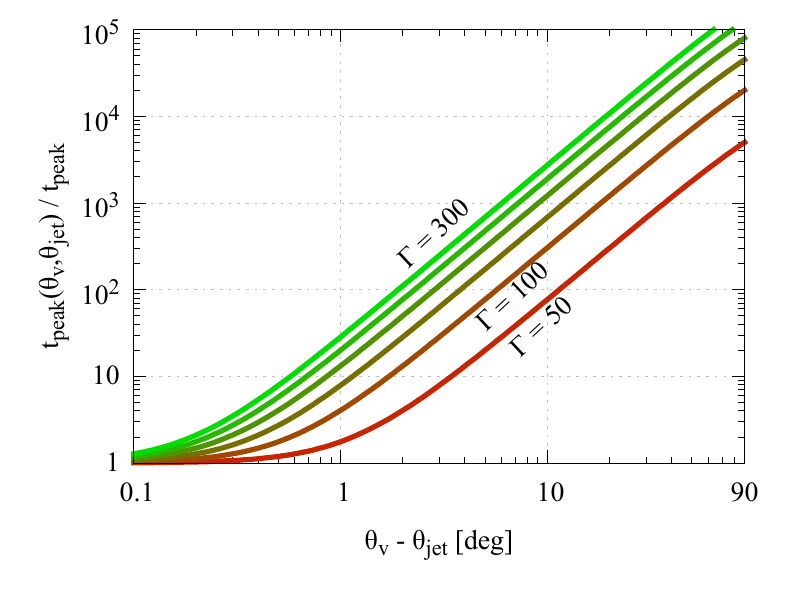} 
 \end{center}
 \caption{\label{fig:tpeak}  Ratio of the off-axis pulse peak time $t_{\rm peak}(\tv,\tj)$ to the on-axis pulse peak time $t_{\rm peak}$. The jet half opening angle is $\tj=5^\circ$. Each curve refers to a different value of the Lorentz factor, from $\Gamma=50$ to $\Gamma = 300$ with a step of $50$.}
\end{figure}
 
 If the jet is off-axis, relativistic beaming of the emitted radiation causes both the flux and $\Ep$ to be much lower than the on-axis counterparts. For the same reason, the duration of the pulse becomes longer. This can be understood intuitively as follows: as in the on-axis case, the jet surface is not seen to turn on all at the same time, but progressively from the nearest-to-the-observer point (point $A$ in Fig.~\ref{fig:geom6}) down to the farthest. The same holds for the turning off. Thus point $A$ is the first to turn on, and also the first to turn off. As a consequence, the effective emitting area increases as long as point $A$ is seen emitting, then it decreases. In other words, the peak time equals the emission time of point $A$, which is given by
\begin{equation}
 t_{\rm peak}(\tv,\tj) = T\left[1-\beta\cos(\tv-\tj)\right]
\end{equation}
thus its ratio to the on-axis peak time is
\begin{equation}
 \frac{t_{\rm peak}(\tv,\tj)}{t_{\rm peak}} = \frac{1-\beta\cos(\tv-\tj)}{1-\beta}
\end{equation} 
Figure~\ref{fig:tpeak} shows a plot of this ratio as a function of $\tv-\tj$ for different values of $\Gamma$. The off-axis pulse is thus intrinsically broader than its on-axis counterpart. The effective duration as seen by the observer, though, depends on the limiting flux and on the amount of overlap with other pulses.

\subsubsection{A lower peak flux}\label{sec:lower-flux}

\begin{figure}
 \begin{center}
 \includegraphics[width=\columnwidth]{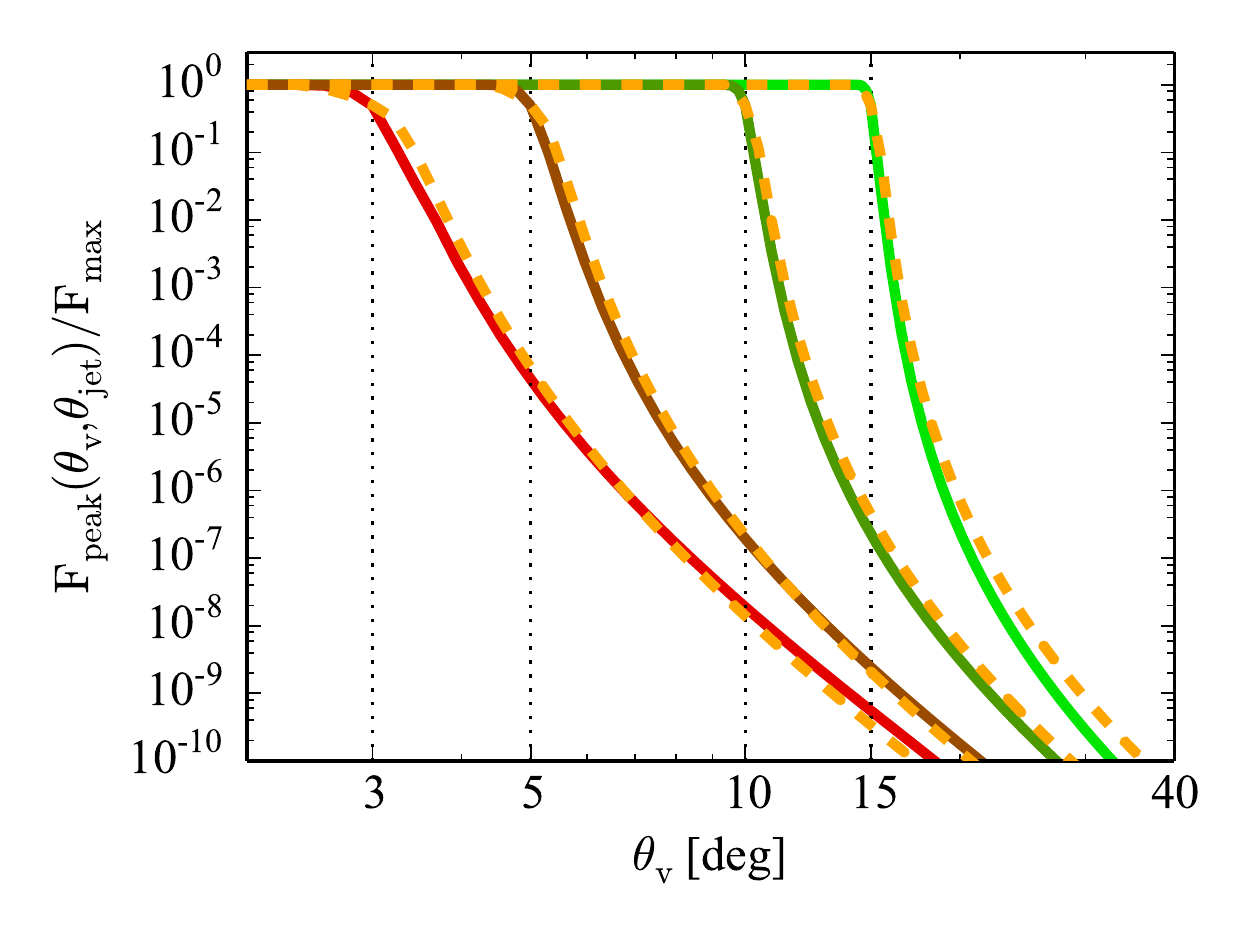} 
 \end{center}
 \caption{\label{fig:peakflux}  Peak fluxes of pulses from jets with four different half opening angles, namely $\tj = 3^\circ$, $5^\circ$, $10^\circ$, and $15^\circ$ (indicated by the thin vertical dotted lines), assuming $\Gamma = 100$ and $\Delta R = R$. The orange dashed curves represent the corresponding empirical parametrization given in Eq.~\ref{eq:peakflux-empirical}.}
\end{figure}

The decrease of the pulse peak flux $F_{\rm p}$ with increasing viewing angle can be understood as follows:
\begin{enumerate}
 \item when the jet is observed on axis, the bulk of the flux comes from a ring of angular radius $1/\Gamma$ centred on the line of sight. Let us indicate this peak flux with $F^{*}$;
 \item as long as $\tv < \tj - 1/\Gamma$, we have that $F_{\rm p}$ is essentially equal to $F^{*}$;
 \item if $\tv = \tj$, about half of the ring is still visible, thus $F_{\rm p}$ is reduced to about $F^{*}/2$;
 \item if $\tv$ is only slightly larger than $\tj$, the flux is dominated by the contribution of the jet border, whose Doppler factor is $\delta_{\rm B} = \Gamma^{-1}\left[1-\beta\cos\left(\tv-\tj\right)\right]^{-1}$, thus $F_{\rm p} \propto \delta_B^4\,F^{*}$;
 \item as $\tv$ increases towards $\tv \gg \tj$, the relative difference in Doppler factor between different parts of the jet is reduced, and the flux contributions of parts other than the border become increasingly important. This compensates in part the de-beaming of the jet border, the effect being more pronounced for larger jets, because the effective emitting surface area is larger.
\end{enumerate}
 Based on these considerations, an empirical analytical formula can be constructed to describe how the peak flux depends on the viewing angle $\tv$ and on the jet half opening angle $\tj$. An example of such an empirical formula is
 \begin{equation}
  F_{\rm p}/F^* \approx \left\lbrace\begin{array}{lr}
                            1 & \tv \leq \tj^* \vspace{12pt}\\
                            1 - \Gamma(\tv - \tj^*)/2 & \tj^* < \tv \leq \tj \vspace{4pt}\\
                            \dfrac{1}{2}\left(\dfrac{\delta_B}{(1+\beta)\Gamma}\right)^{(4 - \sqrt{2}\tj^{1/3})} & \tv > \tj
                           \end{array}\right.\label{eq:peakflux-empirical}
 \end{equation} 
 where $\tj^* = \tj - 1/\Gamma$. The definition for $\tj^* < \tv \leq \tj$ is just a linear decrease from $F^*$ to $F^*/2$; the exponent of $\delta_B$ in the definition for $\tv>\tj$ is $4$ reduced by an amount\footnote{The coefficient and exponent of $\tj$ in Eq.~\ref{eq:peakflux-empirical} have been chosen to get a good agreement with the results from the semi-analytical formulation developed in the Appendix.} which depends on $\tj$, in order to take into account the flux loss compensation explained in point (v) above.
 
 In the Appendix (\S\ref{sec:offaxis-appendix}) we show that the flux at time $t$ of the pulse from an off-axis jet is given by the integral in Eq.~\ref{eq:flux-offaxis}, which however has no analytical solution for $\tv>0$.  The coloured solid lines in Figure~\ref{fig:peakflux} represent $F_{\rm p}$ as computed by numerical integration of Eq.~\ref{eq:flux-offaxis} at $t=t_{\rm peak}(\tv,\tj)$, for five jets with different half opening angles. The orange dashed lines are plots of Eq.~\ref{eq:peakflux-empirical} for the corresponding parameter values, showing that the best agreement is for half opening angles $5^\circ \lesssim \tj \lesssim 10^\circ$.

\subsubsection{Spectral peak energy, hardness-intensity correlation}\label{sec:lower-epeak}

\begin{figure}
 \begin{center}
 \includegraphics[width=\columnwidth]{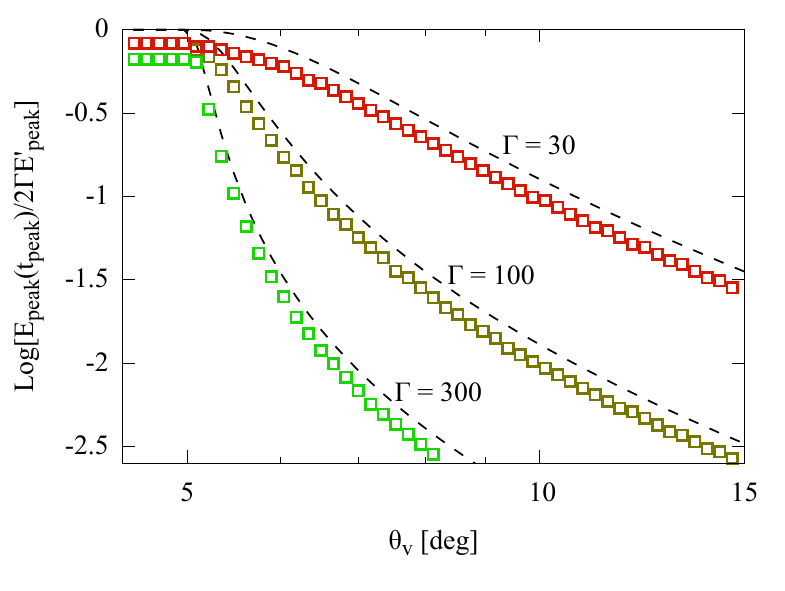} 
 \end{center}
 \caption{\label{fig:epeak-thv} $\Ep$ at the pulse peak time for three jets with $R=10^{13}$ cm, $\tj=5^\circ$ and three values of $\Gamma$, namely (from red to green) $\Gamma = 30$, $100$ and $300$. The black dashed lines are plots of $\delta_B/2\Gamma$ for the corresponding values of $\Gamma$.  }
\end{figure}

With the same assumptions as in the on-axis case, we computed the spectra from the off-axis pulse at different times. The spectrum at each time is dominated by the part of the EATS with the strongest beaming. At time $t_{\rm peak}$, such part is the border of the jet nearest to the observer, thus one expects $\Ep(t_{\rm peak})$ to decrease with $\tv$ as the Doppler factor of the jet border, \ie $\Ep(t_{\rm peak})\propto \delta_B$. Figure~\ref{fig:epeak-thv} is a plot of $\Ep(t_{\rm peak})$ for three values of $\Gamma$, obtained by using Eq.~\ref{eq:off-axis-spectrum} to compute the spectra, and it shows that indeed $\Ep$ is approximately proportional to $\delta_B$. In general, $\Ep$ is a little lower than $\delta_B \Ep'$ because of the ``blending in'' of softer spectra from less beamed parts of the jet.      

\begin{figure}
 \begin{center}
 \includegraphics[width=\columnwidth]{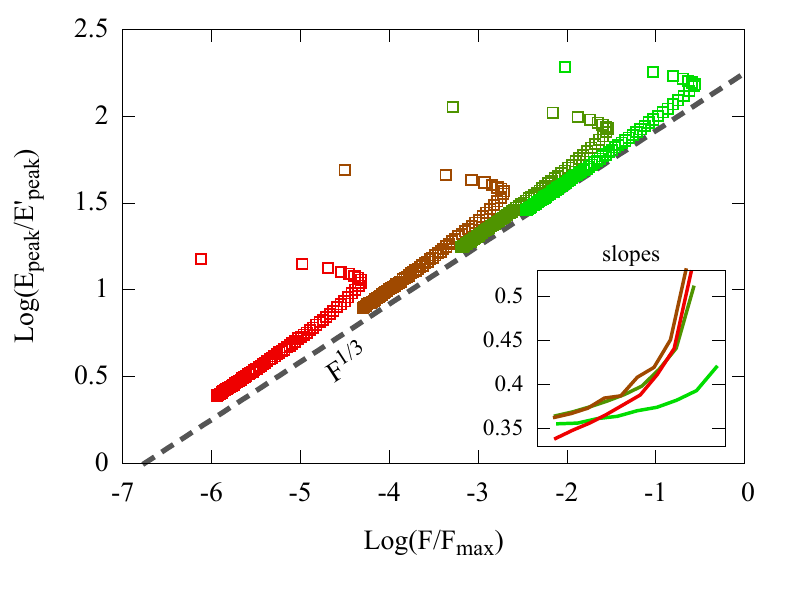} 
 \end{center}
 \caption{\label{fig:epeak-off-axis} Logarithmic plot of $\Ep$ versus Flux of the same pulse seen at different off-axis viewing angles. The jet has $\tj = 5^\circ$, $\Gamma = 100$ and $R=10^{13}$ cm. The four series of points (from green to red) correspond to $\tv = 5.1^\circ$, $5.5^\circ$, $6^\circ$ and $7^\circ$. The inset shows the slope of the relation during the decay of the pulse for each viewing angle.  }
\end{figure}

Figure~\ref{fig:epeak-off-axis} shows the evolution of $\Ep$ as a function of the flux $F$ during the pulse, for four different off-axis viewing angles. A ``hardness-intensity'' correlation during the pulse decay is still apparent, with a slightly steeper slope ($\sim 0.5$) just after the pulse peak, getting shallower as the flux decreases and eventually reaching $\sim 1/3$  as in the on-axis case. 

\subsubsection{Light curves}

\begin{figure}
 \begin{center}
 \includegraphics[width=\columnwidth]{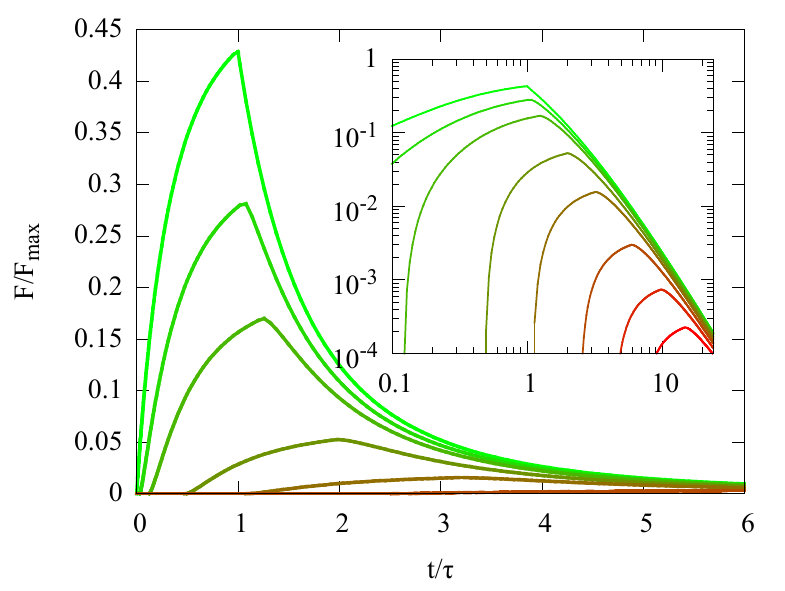} 
 \end{center}
 \caption{\label{fig:homogeneous-off-axis-lcs}  Light curves of a pulse from a jet with $\tj = 5^o$ and $\Gamma = 100$. Each curve refers to a different viewing angle in the sequence (from green to red) $\tv = 5^o$, $5.2^o$,  $5.4^o$, $5.6^o$, $5.8^o$ and $6^o$.  $F_{\rm max}$ refers to the on-axis jet. The inset shows the same curves plotted with logarithmic axes.}
\end{figure}

Figure~\ref{fig:homogeneous-off-axis-lcs} shows plots of bolometric light curves of the same pulse seen at different viewing angles, computed by numerical integration of Eq.~\ref{eq:flux-offaxis} using a Runge-Kutta IV order scheme. Both the peak flux decrease and the duration increase discussed in \S\ref{sec:longer-duration} and \S\ref{sec:lower-flux} are apparent. The overall shape is qualitatively insensitive of the viewing angle, apart from the peak being sharper in the on-axis case.

\section{Multi-pulse light curves}\label{sec:multipulse}

\begin{figure*}
 \begin{center}
 \includegraphics[width=\textwidth]{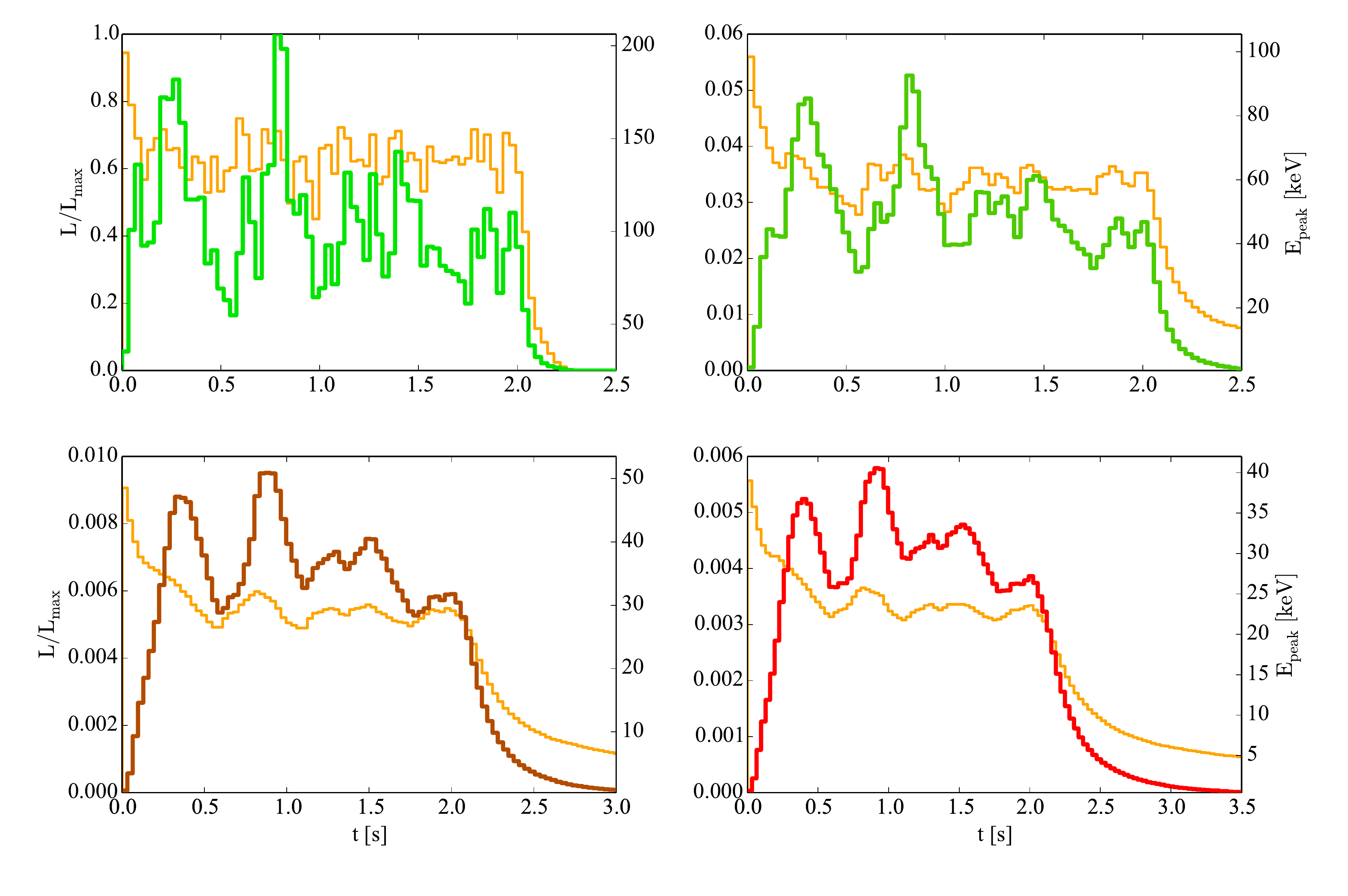} 
 \end{center}
 \caption{\label{fig:offaxis-multi-lcs}  Light curves (thick histograms) and spectral peak evolution (thin orange histograms) of a sequence of $100$ pulses from a jet with $\tj = 5^o$, $\Gamma = 100$, $R=10^{13}$ cm and $\Delta R = R$. The peak of the $\nu F_\nu$ comoving spectrum is $E_{\rm p}'=1$ keV. The pulse start times are randomly distributed within the first $2$ s of the observer time. Each panel refers to a different viewing angle in the sequence $\tv = 0$, $\tj+1/\Gamma$, $\tj+\sqrt{3}/\Gamma$, $\tj+2/\Gamma$ (from left to right, top to bottom). $L_{\rm max}$ refers to the peak luminosity of the on-axis light curve.}
\end{figure*}

Now that we have a detailed (though simple) model of the single pulse, we can proceed to construct a ``synthetic'' GRB light curve by superposition of pulses.
Some non-trivial features emerge from such superposition. Figure~\ref{fig:offaxis-multi-lcs} shows four light curves of the same series of $N=100$ pulses seen at four different viewing angles. All pulses are equal in duration and peak flux. Their starting times have been sampled from a uniform distribution within a $2$ seconds time span. The jet parameters are $\Gamma=100$, $\tj=5^\circ$, $R=10^{13}$ cm and $\Delta R = R$. The viewing angles are $\tv=0$, $\tj+1/\Gamma$, $\tj+\sqrt{3}/\Gamma$ and $\tj+2/\Gamma$. The comoving spectral shape is the same as before (Eq.~\ref{eq:comov-spec-shape}). The resulting light curves have been binned at $32$ ms resolution for a better comparison with actual GRB light curves. For each light curve, the $\Ep$ of the spectrum in each time bin is also given (thin orange histograms). The following features should be apparent:
\begin{enumerate}
 \item as the viewing angle increases, variability is smeared out by the pulse broadening;
 \item the shape of the overall light curve tends to resemble a (long) single pulse when the viewing angle is large enough;
 \item the superposition of pulses masks the hard-to-soft spectral evolution of the single pulses, turning it into an intensity tracking behaviour: this is due to the superposition of spectra with different peak energies;
 \item the variation of $\Ep$ leads slightly the variation in flux, because of the hard-to-soft nature of the single pulses;
 \item there is a general softening of $\Ep$ in time over the entire light curve.
\end{enumerate}
These features are strikingly similar to those found in time resolved spectral analysis of real \GRBs \citep[\eg][]{Ford1995,Ghirlanda2002}. We do not advocate this as a proof of the correctness of our model, which is certainly oversimplified, but rather as a further indication that some features of GRB light curves might be explained admitting that the jet is seen at least slightly off-axis. The off-axis viewing angle favours the broadening and superposition of pulses, which is the necessary ingredient to some of the features enumerated above. It can also contribute in a simple way to explain why the slope of the hardness-intensity correlation changes from burst to burst, being influenced by the viewing angle (\S\ref{sec:lower-epeak}).

Figure~\ref{fig:offaxis-multi-lcs} shows that the simple arguments outlined in \S\ref{sec:naive} are valid not only if pulses are produced by point sources, but also in presence of an extended geometry.


\section{The number of off-axis GRBs}\label{sec:n_offaxis}
We can obtain an estimate of the number of off-axis GRBs over the observed population by the simplifying assumption that all jets share the same intrinsic properties, and that their flux in an observer band is uniquely determined by the viewing angle and the redshift. We assume that the majority of GRBs are observed on-axis, and we choose the following parameters in an attempt to match the average properties of the on-axis population:
\begin{enumerate}
 \item $E_{\rm peak,0}=560\, \rm{keV}$ as the typical (on-axis, rest frame) peak spectral energy, chosen to match the average value of the \textit{Fermi}/GBM sample $\left<E_{\rm peak}^{\rm obs}\right>\sim 186\, \rm{keV}$ \citep{2011A&A...530A..21N} multiplied by a typical redshift $\left<1+z\right>\sim 3$;
 \item $\alpha = -0.86$ and $\beta = -2.3$ as typical low- and high-energy spectral indices \citep{2011A&A...530A..21N};
 \item $\Psi(z) = (0.0157+0.118 z)/(1+(z/3.23)^{4.66})$ as the redshift distribution, \ie the GRB formation rate as given in \cite{Ghirlanda2015high_redshift};
 \item $L_0=2.5\times 10^{52} \rm{erg\,s^{-1}}$ as the typical (on-axis) luminosity, which corresponds to the break of the broken power law luminosity function (model with no redshift evolution) of the BAT6 complete sample \citep{2012ApJ...749...68S}. This choice is motivated by the fact that if GRBs can be observed off-axis, then their luminosity function is indeed well described by a broken power law, with the break around the average on-axis luminosity \citep{2015MNRAS.447.1911P};
 \item since the result is sensitive to the assumed typical Lorentz factor $\Gamma$ and half-opening angle $\tj$, we explore the cases $\Gamma=50$, 100 and 300, and $\tj=5^\circ$ and $10^\circ$.
\end{enumerate}

We then define the effective luminosity $L(\tv)$ following Eq.~\ref{eq:peakflux-empirical}, namely
\begin{equation}
 L(\tv) = L_0\times\left\lbrace\begin{array}{lr}
                                1 & \tv \leq \tj^*\\
                                \vspace{5pt}
                                1-\Gamma(\tv-\tj^*)/2 & \tj^* \leq \tv < \tj\\
                                \vspace{5pt}
                                \dfrac{1}{2}\left(\dfrac{\delta_B}{(1+\beta)\Gamma}\right)^{\left(4-\sqrt{2}\tj^{1/3}\right)} & \tv>\tj\\
                               \end{array}\right.\label{eq:Lthv}
\end{equation} 
with $\tj^* = \tj-\Gamma^{-1}$, and the effective peak energy 
\begin{equation}
 E_{\rm peak}(\tv)=\frac{E_{\rm peak,0}}{1+z}\times\left\lbrace\begin{array}{lr}
                                                    1 & \tv \leq \tj\\
                                                    \frac{\delta_B}{(1+\beta)\Gamma} & \tv > \tj\\
                                                   \end{array}\right.
\end{equation} 
as in \S\ref{sec:lower-epeak}. With these assumptions and prescriptions, we can compute the observed rate of GRBs with a viewing angle in the range $(\tv,\tv+d\tv)$, in the redshift range $(z,z+dz)$, assuming a limiting photon flux $p_{\rm lim}$ in a given band, as
\begin{equation}
 \dfrac{d\dot{N}}{d\tv\,dz}d\tv\,dz = \dfrac{\Psi(z)}{1+z}\dfrac{dV}{dz} P(\tv,z,p_{\rm lim})d\tv\,dz \label{eq:dNdthvdz}
\end{equation} 
where $P(\tv,z,p_{\rm lim})$ is the viewing angle probability, $dV/dz$ is the differential comoving volume, and the factor $1+z$ accounts for cosmological time dilation.
The viewing angle probability is
\begin{equation}
P(\tv,z,p_{\rm lim}) = \left\lbrace\begin{array}{lr}
                      \sin\tv & \tv\leq\theta_{\rm v,lim}(z,p_{\rm lim})\\
                      0 & \tv>\theta_{\rm v,lim}(z,p_{\rm lim})\\
                     \end{array}\right.
\end{equation} 
The limiting viewing angle $\theta_{\rm{v,lim}}$ corresponds (through Eq.~\ref{eq:Lthv}) to the limiting luminosity $L_{\rm{lim}}$ computed as
\begin{equation}
 L_{\rm{lim}} = 4\pi d_L^2\,p_{\rm{lim}} \frac{\int_0^\infty \frac{dN}{dE}E\,dE}{\int_{(1+z)E_{\rm{low}}}^{(1+z)E_{\rm{high}}}\frac{dN}{dE}\,dE}
\end{equation}
where $E_{\rm{low}}$ ($E_{\rm{high}}$) is the lower (upper) limit of the observer band, $d_L$ is the luminosity distance, and $dN/dE(\Ep,\alpha,\beta)$ is the rest frame spectrum.

We define the total rate $\dot{N}_{\rm tot}(<z)$  of observable GRBs within redshift $z$ as the integral of Eq.~\ref{eq:dNdthvdz} over redshift from $0$ to $z$ and over $\tv$ from $0$ to $\pi/2$; similarly, the rate $\dot{N}_{\rm off}(<z)$ of off-axis GRBs within redshift $z$ is the integral over redshift from $0$ to $z$ and over the viewing angle from $\tv+\Gamma^{-1}$ to $\pi/2$. Since we are interested in the ratio of these two quantities, we do not need to bother about the normalization.

\begin{figure}
 \includegraphics[width=\columnwidth]{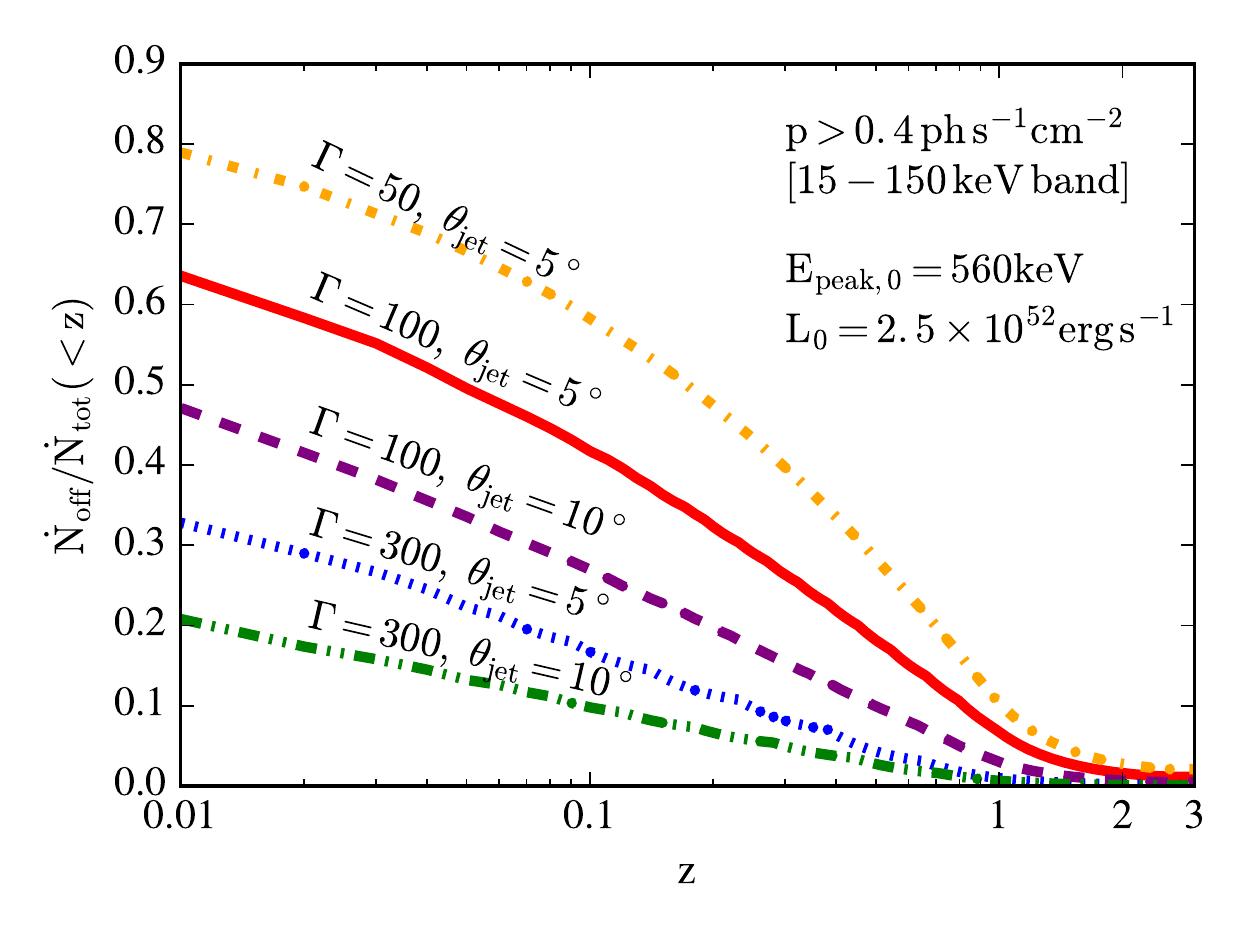}
 \caption{\label{fig:n_offaxis}Fraction of off-axis GRBs over the total within a given redshift. The curves represent an estimate of the fraction of GRBs with redshift lower than $z$ observable by \textit{Swift}/BAT (i.e.\ with photon flux $p>0.4\,\rm{ph\,s^{-1}cm^{-2}}$ in the 15--150 keV band) whose viewing angle is larger than $\tj+\Gamma^{-1}$.}
\end{figure}

In Fig.~\ref{fig:n_offaxis} we show the fraction of bursts with $\tv>\tj+\Gamma^{-1}$ at redshift lower than $z$ for various choices of $\Gamma$ and $\tj$, assuming a limiting flux $p_{\rm{lim}}=0.4\,\rm{ph\,s^{-1}\,cm^{-2}}$ in the $15$-$150\,\rm{keV}$ band, to reproduce the \textit{Swift}/BAT band and limiting flux. Standard flat $\Lambda$CDM cosmology was assumed, with Planck parameters $H_0 = 67.3\,\rm{km\,s^{-1}\,Mpc^{-1}}$ and $\Omega_{m,0} = 0.315$ \citep{Planck2014cosmoparams}. These results clearly indicate that at low redshift a significant fraction of GRBs is likely seen off-axis. 

\subsection{Low luminosity GRBs}\label{sec:low-lum-grbs}
Recently, some authors \citep[\eg][]{Liang2007,Zhang2008,He2009,2011ApJ...739L..55B} argued that a unique population of low luminosity \GRBs exists, based on some common features of GRB060218, GRB980425, GRB031203 and GRB100316D. These features include, apart from the low inferred isotropic equivalent luminosity, an apparently single pulsed, smooth light curve (low variability), and a low average $\Ep$. Since all such bursts were at a low redshift ($z\lesssim 0.1$), the rate of like events per unit comoving volume in the Universe is very high (higher than the rate of ``normal'' GRBs), but we do not see the majority of these events because of selection effects.
The results discussed in this paper suggest that the apparently peculiar features of these GRBs can be interpreted instead as being just the indication that they were observed off-axis. Moreover, in \cite{2015MNRAS.447.1911P} we have shown that the observed rate of low luminosity GRBs is consistent with what one would expect if they were just ordinary bursts seen off-axis. Based on these considerations, we argue that there is no need to invoke a new separate population of low luminosity GRBs.

\section{Discussion and Conclusions}\label{sec:conclusions}

In this work we set up a simple physical model of a single GRB pulse based on shell curvature only, as a tool to explore the effect of the viewing angle on GRB light curves. Compared to other similar (and more refined) models \citep[\eg][]{Dermer2004,Genet2009}, our model includes the effect of an off-axis viewing angle. We show that the inclusion of such effect is important because a significant fraction (from 10\% up to 80\%) of nearby bursts ($z<0.1$) are likely observed off-axis. Admittedly, the assumptions behind the pulse model are at best a very rough approximation of reality. The general trend of the effect of the viewing angle, though, is largely insensitive of the simplifications adopted: a slightly off-axis viewing angle is enough to produce a significant pulse broadening, without affecting the pulse separation. This in turn leads to pulse overlap, which smears out variability at all frequencies, resulting in a smoother light curve and spectral evolution.

This is mainly a consequence of two assumptions: (i) that the emission is isotropic in the comoving frame and (ii) that all pulses are produced around a typical radius. By relaxing (i), \ie allowing for a strongly anisotropic emission in the comoving frame, one could reduce (in case the anisotropy favours forward emission) or enhance (in case the anisotropy favours backwards emission) the flux received by off-axis observers. By relaxing (ii), on the other hand, one may have that the pulse separation depends on the viewing angle as well. One would then need to explain, though, why the observed pulse width distribution does not vary in time \citep{Ramirez-Ruiz1999,piran-physics_of_grbs04}, despite the change of the emission radius.

Given the above considerations, we conclude that:

    \begin{enumerate}
    \item if the GRB jet is seen off-axis, single pulses appear longer and their spectrum appears softer than in the on-axis case;
    \item if a burst is made up of a superposition of pulses, its variability is smeared out by pulse broadening if the jet is observed off-axis, with respect to the on-axis case;
    \item if single pulses feature an intrinsic hard-to-soft spectral evolution, pulse overlap can turn it into an intensity-tracking behaviour.
    \end{enumerate}


As discussed in \S\ref{sec:low-lum-grbs}, the results support the idea that prompt emission properties of so-called low luminosity GRBs can be interpreted as indications that they are just ordinary bursts seen off-axis.

\section*{Acknowledgements}

O. S. Salafia thanks the \textit{Swift} group at the Astronomical Observatory of Brera - Merate for useful discussions, Ryo Yamazaki for kindly pointing out some important references which had not been cited in the first draft, and the anonymous referee for insightful comments which stimulated a significant improvement of this work.
 
\section*{References}

\footnotesize{
\bibliographystyle{mnras}
\bibliography{journals,bibtex}

\begin{thebibliography}{}
\makeatletter
\relax
\def\mn@urlcharsother{\let\do\@makeother \do\$\do\&\do\#\do\^\do\_\do\%\do\~}
\def\mn@doi{\begingroup\mn@urlcharsother \@ifnextchar [ {\mn@doi@}
  {\mn@doi@[]}}
\def\mn@doi@[#1]#2{\def\@tempa{#1}\ifx\@tempa\@empty \href
  {http://dx.doi.org/#2} {doi:#2}\else \href {http://dx.doi.org/#2} {#1}\fi
  \endgroup}
\def\mn@eprint#1#2{\mn@eprint@#1:#2::\@nil}
\def\mn@eprint@arXiv#1{\href {http://arxiv.org/abs/#1} {{\tt arXiv:#1}}}
\def\mn@eprint@dblp#1{\href {http://dblp.uni-trier.de/rec/bibtex/#1.xml}
  {dblp:#1}}
\def\mn@eprint@#1:#2:#3:#4\@nil{\def\@tempa {#1}\def\@tempb {#2}\def\@tempc
  {#3}\ifx \@tempc \@empty \let \@tempc \@tempb \let \@tempb \@tempa \fi \ifx
  \@tempb \@empty \def\@tempb {arXiv}\fi \@ifundefined
  {mn@eprint@\@tempb}{\@tempb:\@tempc}{\expandafter \expandafter \csname
  mn@eprint@\@tempb\endcsname \expandafter{\@tempc}}}

\bibitem[\protect\citeauthoryear{Basak \& Rao}{Basak \& Rao}{2014}]{Basak2014}
Basak R.,  Rao A.~R.,  2014, \mn@doi [Monthly Notices of the Royal Astronomical
  Society] {10.1093/mnras/stu882}, 442, 419

\bibitem[\protect\citeauthoryear{Beloborodov}{Beloborodov}{2010}]{Beloborodov2010}
Beloborodov A.~M.,  2010, \mn@doi [Monthly Notices of the Royal Astronomical
  Society] {10.1111/j.1365-2966.2010.16770.x}, 407, 1033

\bibitem[\protect\citeauthoryear{{Borgonovo} \& {Ryde}}{{Borgonovo} \&
  {Ryde}}{2001}]{Borgonovo2001ApJ}
{Borgonovo} L.,  {Ryde} F.,  2001, \mn@doi [\apj] {10.1086/319008}, \href
  {http://adsabs.harvard.edu/abs/2001ApJ...548..770B} {548, 770}

\bibitem[\protect\citeauthoryear{Bromberg, Nakar  \& Piran}{Bromberg
  et~al.}{2011}]{2011ApJ...739L..55B}
Bromberg O.,  Nakar E.,   Piran T.,  2011, \mn@doi [The Astrophysical Journal]
  {10.1088/2041-8205/739/2/L55}, 739, L55

\bibitem[\protect\citeauthoryear{Daigne \& Mochkovitch}{Daigne \&
  Mochkovitch}{2002}]{daigne-thermal-precursors-2002}
Daigne F.,  Mochkovitch R.,  2002, \mn@doi [Monthly Notices of the Royal
  Astronomical Society] {10.1046/j.1365-8711.2002.05875.x}, 336, 1271

\bibitem[\protect\citeauthoryear{Dermer}{Dermer}{2004}]{Dermer2004}
Dermer C.~D.,  2004, \mn@doi [The Astrophysical Journal] {10.1086/426532}, 614,
  284

\bibitem[\protect\citeauthoryear{Ford et~al.,}{Ford et~al.}{1995}]{Ford1995}
Ford L.~A.,  et~al., 1995, \mn@doi [Astrophysical Journal] {10.1086/175174},
  439, 307

\bibitem[\protect\citeauthoryear{Genet \& Granot}{Genet \&
  Granot}{2009}]{Genet2009}
Genet F.,  Granot J.,  2009, \mn@doi [Monthly Notices of the Royal Astronomical
  Society] {10.1111/j.1365-2966.2009.15355.x}, 399, 1328

\bibitem[\protect\citeauthoryear{Ghirlanda, Celotti  \& Ghisellini}{Ghirlanda
  et~al.}{2002}]{Ghirlanda2002}
Ghirlanda G.,  Celotti A.,   Ghisellini G.,  2002, \mn@doi [Astronomy and
  Astrophysics] {10.1051/0004-6361:20021038}, 393, 409

\bibitem[\protect\citeauthoryear{Ghirlanda, Salvaterra, Ghisellini, Mereghetti,
  Tagliaferri  \& Others}{Ghirlanda et~al.}{2015}]{Ghirlanda2015high_redshift}
Ghirlanda G.,  Salvaterra R.,  Ghisellini G.,  Mereghetti S.,  Tagliaferri G.,
   Others 2015, \mn@doi [Monthly Notices of the Royal Astronomical Society]
  {10.1093/mnras/stv183}

\bibitem[\protect\citeauthoryear{Ghisellini}{Ghisellini}{2013}]{ghisellini_book_2013}
Ghisellini G.,  2013, {Radiative Processes in High Energy Astrophysics}, 1 edn.
 Lecture Notes in Physics, Berlin Springer Verlag Vol. 873, Springer
  International Publishing (\mn@eprint {arXiv} {1202.5949}),
  \mn@doi{10.1007/978-3-319-00612-3}

\bibitem[\protect\citeauthoryear{Ghisellini, Celotti  \& Lazzati}{Ghisellini
  et~al.}{2000}]{Ghisellini2000}
Ghisellini G.,  Celotti A.,   Lazzati D.,  2000, \mn@doi [Monthly Notices of
  the Royal Astronomical Society] {10.1046/j.1365-8711.2000.03354.x}, 313, L1

\bibitem[\protect\citeauthoryear{Ghisellini, Ghirlanda, Mereghetti, Bosnjak,
  Tavecchio  \& Firmani}{Ghisellini et~al.}{2006}]{2006MNRAS.372.1699G}
Ghisellini G.,  Ghirlanda G.,  Mereghetti S.,  Bosnjak Z.,  Tavecchio F.,
  Firmani C.,  2006, \mn@doi [Monthly Notices of the Royal Astronomical
  Society] {10.1111/j.1365-2966.2006.10972.x}, 372, 1699

\bibitem[\protect\citeauthoryear{Giannios}{Giannios}{2006}]{Giannios2006}
Giannios D.,  2006, \mn@doi [Astronomy and Astrophysics]
  {10.1051/0004-6361:20065000}, 457, 763

\bibitem[\protect\citeauthoryear{{Golenetskii}, {Mazets}, {Aptekar}  \&
  {Ilinskii}}{{Golenetskii} et~al.}{1983}]{Golenetskii1983Nature}
{Golenetskii} S.~V.,  {Mazets} E.~P.,  {Aptekar} R.~L.,   {Ilinskii} V.~N.,
  1983, \mn@doi [Nature] {10.1038/306451a0}, \href
  {http://adsabs.harvard.edu/abs/1983Natur.306..451G} {306, 451}

\bibitem[\protect\citeauthoryear{Granot, Ramirez-Ruiz  \& Perna}{Granot
  et~al.}{2005}]{Granot2005a}
Granot J.,  Ramirez-Ruiz E.,   Perna R.,  2005, \mn@doi [The Astrophysical
  Journal] {10.1086/431477}, 630, 1003

\bibitem[\protect\citeauthoryear{Hakkila \& Preece}{Hakkila \&
  Preece}{2011}]{2011ApJ...740..104H}
Hakkila J.,  Preece R.,  2011, \mn@doi [Astrophys.J.]
  {10.1088/0004-637X/740/2/104}, 740, 104

\bibitem[\protect\citeauthoryear{He, Wang, Yu  \& M{\'{e}}sz{\'{a}}ros}{He
  et~al.}{2009}]{He2009}
He H.-N.,  Wang X.-Y.,  Yu Y.-W.,   M{\'{e}}sz{\'{a}}ros P.,  2009, \mn@doi
  [The Astrophysical Journal] {10.1088/0004-637X/706/2/1152}, 706, 1152

\bibitem[\protect\citeauthoryear{Imhof, Nakano, Johnson, Kilner, Regan,
  Klebesadel  \& Strong}{Imhof et~al.}{1974}]{Imhof1974}
Imhof W.~L.,  Nakano G.~H.,  Johnson R.~G.,  Kilner J.~R.,  Regan J.~B.,
  Klebesadel R.~W.,   Strong I.~B.,  1974, \mn@doi [The Astrophysical Journal]
  {10.1086/181529}, 191, L7

\bibitem[\protect\citeauthoryear{Ioka \& Nakamura}{Ioka \&
  Nakamura}{2001}]{Ioka2001}
Ioka K.,  Nakamura T.,  2001, \mn@doi [The Astrophysical Journal]
  {10.1086/321717}, 554, L163

\bibitem[\protect\citeauthoryear{Kargatis \& Liang}{Kargatis \&
  Liang}{1995}]{Kargatis1995}
Kargatis V.~E.,  Liang E.~P.,  1995, \mn@doi [Astrophysics and Space Science]
  {10.1007/BF00658611}, 231, 177

\bibitem[\protect\citeauthoryear{{Kargatis}, {Liang}, {Hurley}, {Barat},
  {Eveno}  \& {Niel}}{{Kargatis} et~al.}{1994}]{Kargatis1994ApJ}
{Kargatis} V.~E.,  {Liang} E.~P.,  {Hurley} K.~C.,  {Barat} C.,  {Eveno} E.,
  {Niel} M.,  1994, \mn@doi [\apj] {10.1086/173724}, \href
  {http://adsabs.harvard.edu/abs/1994ApJ...422..260K} {422, 260}

\bibitem[\protect\citeauthoryear{Kumar \& Panaitescu}{Kumar \&
  Panaitescu}{2000}]{Kumar2000}
Kumar P.,  Panaitescu A.,  2000, \mn@doi [The Astrophysical Journal]
  {10.1086/312905}, 541, L51

\bibitem[\protect\citeauthoryear{Lazarian, Petrosian, Yan  \& Cho}{Lazarian
  et~al.}{2003}]{Lazarian2003}
Lazarian a.,  Petrosian V.,  Yan H.,   Cho J.,  2003, eprint
  arXiv:astro-ph/0301181, p.~18

\bibitem[\protect\citeauthoryear{Lazzati, Ghisellini  \& Celotti}{Lazzati
  et~al.}{1999}]{Lazzati1999}
Lazzati D.,  Ghisellini G.,   Celotti A.,  1999, \mn@doi [Monthly Notices of
  the Royal Astronomical Society] {10.1046/j.1365-8711.1999.02970.x}, 309, L13

\bibitem[\protect\citeauthoryear{Lee, Bloom  \& Petrosian}{Lee
  et~al.}{2000}]{Lee2000}
Lee A.,  Bloom E.~D.,   Petrosian V.,  2000, \mn@doi [The Astrophysical Journal
  Supplement Series] {10.1086/317364}, 131, 1

\bibitem[\protect\citeauthoryear{Liang \& Kargatis}{Liang \&
  Kargatis}{1996}]{Liang1996}
Liang E.,  Kargatis V.,  1996, \mn@doi [Nature] {10.1038/381049a0}, 381, 49

\bibitem[\protect\citeauthoryear{Liang, Zhang  \& Zhang}{Liang
  et~al.}{2007}]{Liang2007}
Liang E.,  Zhang B.,   Zhang B.,  2007, \mn@doi [The Astrophysical Journal]
  {10.1086/521870}, 670, 565

\bibitem[\protect\citeauthoryear{Link, Epstein  \& Priedhorsky}{Link
  et~al.}{1993}]{Link1993}
Link B.,  Epstein R.~I.,   Priedhorsky W.~C.,  1993, \mn@doi [The Astrophysical
  Journal] {10.1086/186836}, 408, L81

\bibitem[\protect\citeauthoryear{Lu, Wei, Liang, Zhang, L{\"{u}}, L{\"{u}}, Lei
   \& Zhang}{Lu et~al.}{2012}]{Lu2012}
Lu R.-J.,  Wei J.-J.,  Liang E.-W.,  Zhang B.-B.,  L{\"{u}} H.-J.,  L{\"{u}}
  L.-Z.,  Lei W.-H.,   Zhang B.,  2012, \mn@doi [The Astrophysical Journal]
  {10.1088/0004-637X/756/2/112}, 756, 112

\bibitem[\protect\citeauthoryear{Nakar}{Nakar}{2015}]{Nakar2015}
Nakar E.,  2015, \mn@doi [The Astrophysical Journal]
  {10.1088/0004-637X/807/2/172}, 807, 172

\bibitem[\protect\citeauthoryear{Nakar \& Piran}{Nakar \&
  Piran}{2002a}]{Nakar2002b}
Nakar E.,  Piran T.,  2002a, \mn@doi [Monthly Notices of the Royal Astronomical
  Society] {10.1046/j.1365-8711.2002.05158.x}, 331, 40

\bibitem[\protect\citeauthoryear{Nakar \& Piran}{Nakar \&
  Piran}{2002b}]{Nakar2002a}
Nakar E.,  Piran T.,  2002b, \mn@doi [The Astrophysical Journal]
  {10.1086/341748}, 572, L139

\bibitem[\protect\citeauthoryear{Nava, Ghirlanda, Ghisellini  \& Celotti}{Nava
  et~al.}{2011}]{2011A&A...530A..21N}
Nava L.,  Ghirlanda G.,  Ghisellini G.,   Celotti A.,  2011, \mn@doi [Astronomy
  and Astrophysics] {10.1051/0004-6361/201016270}, 530, A21

\bibitem[\protect\citeauthoryear{Norris, Share, Messina, Dennis, Desai, Cline,
  Matz  \& Chupp}{Norris et~al.}{1986}]{Norris1986}
Norris J.~P.,  Share G.~H.,  Messina D.~C.,  Dennis B.~R.,  Desai U.~D.,  Cline
  T.~L.,  Matz S.~M.,   Chupp E.~L.,  1986, \mn@doi [The Astrophysical Journal]
  {10.1086/163889}, 301, 213

\bibitem[\protect\citeauthoryear{Norris, Nemiroff, Bonnell, Scargle,
  Kouveliotou, Paciesas, Meegan  \& Fishman}{Norris et~al.}{1996}]{Norris1996}
Norris J.~P.,  Nemiroff R.~J.,  Bonnell J.~T.,  Scargle J.~D.,  Kouveliotou C.,
   Paciesas W.~S.,  Meegan C.~A.,   Fishman G.~J.,  1996, \mn@doi [The
  Astrophysical Journal] {10.1086/176902}, 459, 393

\bibitem[\protect\citeauthoryear{Norris, Marani  \& Bonnell}{Norris
  et~al.}{2000}]{Norris2000}
Norris J.~P.,  Marani G.~F.,   Bonnell J.~T.,  2000, \mn@doi [The Astrophysical
  Journal] {10.1086/308725}, 534, 248

\bibitem[\protect\citeauthoryear{Pescalli, Ghirlanda, Salafia, Ghisellini,
  Nappo  \& Others}{Pescalli et~al.}{2015}]{2015MNRAS.447.1911P}
Pescalli A.,  Ghirlanda G.,  Salafia O.~S.,  Ghisellini G.,  Nappo F.,   Others
  2015, \mn@doi [Mon.Not.Roy.Astron.Soc.] {10.1093/mnras/stu2482}, 447, 1911

\bibitem[\protect\citeauthoryear{Piran}{Piran}{2005}]{piran-physics_of_grbs04}
Piran T.,  2005, \mn@doi [Reviews of Modern Physics]
  {10.1103/RevModPhys.76.1143}, 76, 1143

\bibitem[\protect\citeauthoryear{{Planck Collaboration}}{{Planck
  Collaboration}}{2013}]{Planck2014cosmoparams}
{Planck Collaboration} 2013, \mn@doi [Astronomy {\&} Astrophysics]
  {10.1051/0004-6361/201321591}, 571, A16

\bibitem[\protect\citeauthoryear{Preece, Pendleton, Briggs, Mallozzi, Paciesas,
  Band, Matteson  \& Meegan}{Preece et~al.}{1998}]{Preece1998}
Preece R.~D.,  Pendleton G.~N.,  Briggs M.~S.,  Mallozzi R.~S.,  Paciesas
  W.~S.,  Band D.~L.,  Matteson J.~L.,   Meegan C.~A.,  1998, \mn@doi [The
  Astrophysical Journal] {10.1086/305402}, 496, 849

\bibitem[\protect\citeauthoryear{Ramirez-Ruiz \& Fenimore}{Ramirez-Ruiz \&
  Fenimore}{1999}]{Ramirez-Ruiz1999}
Ramirez-Ruiz E.,  Fenimore E.~E.,  1999, \mn@doi [Astronomy and Astrophysics
  Supplement Series] {10.1051/aas:1999336}, 138, 521

\bibitem[\protect\citeauthoryear{Ramirez-Ruiz, Granot, Kouveliotou, Woosley,
  Patel  \& Mazzali}{Ramirez-Ruiz et~al.}{2005}]{Ramirez-Ruiz2005}
Ramirez-Ruiz E.,  Granot J.,  Kouveliotou C.,  Woosley S.~E.,  Patel S.~K.,
  Mazzali P.~A.,  2005, \mn@doi [The Astrophysical Journal] {10.1086/431237},
  625, L91

\bibitem[\protect\citeauthoryear{Rees \& Meszaros}{Rees \&
  Meszaros}{1994}]{Rees1994}
Rees M.~J.,  Meszaros P.,  1994, \mn@doi [Astrophys.J.L.] {10.1086/187446},
  430, L93

\bibitem[\protect\citeauthoryear{Rees \& M{\'{e}}sz{\'{a}}ros}{Rees \&
  M{\'{e}}sz{\'{a}}ros}{2005}]{2005ApJ...628..847R}
Rees M.,  M{\'{e}}sz{\'{a}}ros P.,  2005, \mn@doi [Astrophys.J.]
  {10.1086/430818}, 628, 847

\bibitem[\protect\citeauthoryear{Reichart, Lamb, Fenimore, Ramirez-Ruiz, Cline
  \& Hurley}{Reichart et~al.}{2001}]{Reichart2001}
Reichart D.~E.,  Lamb D.~Q.,  Fenimore E.~E.,  Ramirez-Ruiz E.,  Cline T.~L.,
  Hurley K.,  2001, \mn@doi [The Astrophysical Journal] {10.1086/320434}, 552,
  57

\bibitem[\protect\citeauthoryear{Rhoads}{Rhoads}{1997}]{rhoads-balloon97}
Rhoads J.,  1997, \mn@doi [Astrophys.J.L.] {10.1086/310876}, 487, L1

\bibitem[\protect\citeauthoryear{Rybicki \& Lightman}{Rybicki \&
  Lightman}{1979}]{Rybicki1979}
Rybicki G.~B.,  Lightman A.~P.,  1979, {Radiative Processes in Astrophysics},
  first edn.
 A Wiley-Interscience publication Vol. 25, Wiley-Interscience,
  \mn@doi{10.1016/0031-9201(81)90057-1}, \url
  {http://adsabs.harvard.edu/abs/1979rpa..book.....R}

\bibitem[\protect\citeauthoryear{Ryde \& Petrosian}{Ryde \&
  Petrosian}{2002}]{Ryde2002}
Ryde F.,  Petrosian V.,  2002, \mn@doi [The Astrophysical Journal]
  {10.1086/342271}, 578, 290

\bibitem[\protect\citeauthoryear{Ryde \& Svensson}{Ryde \&
  Svensson}{1998}]{Ryde1998}
Ryde F.,  Svensson R.,  1998, \mn@doi [The Astrophysical Journal]
  {10.1086/306818}, 512, 7

\bibitem[\protect\citeauthoryear{Salvaterra et~al.,}{Salvaterra
  et~al.}{2012}]{2012ApJ...749...68S}
Salvaterra R.,  et~al., 2012, \mn@doi [Astrophys.J.]
  {10.1088/0004-637X/749/1/68}, 749, 68

\bibitem[\protect\citeauthoryear{Soderberg et~al.,}{Soderberg
  et~al.}{2004}]{Soderberg2004}
Soderberg A.~M.,  et~al., 2004, \mn@doi [Nature] {10.1038/nature02757}, 430,
  648

\bibitem[\protect\citeauthoryear{Woods \& Loeb}{Woods \&
  Loeb}{1999}]{1999ApJ...523..187W}
Woods E.,  Loeb A.,  1999, \mn@doi [The Astrophysical Journal]
  {10.1086/307738}, 523, 187

\bibitem[\protect\citeauthoryear{Yamazaki, Ioka  \& Nakamura}{Yamazaki
  et~al.}{2002}]{Yamazaki2002}
Yamazaki R.,  Ioka K.,   Nakamura T.,  2002, \mn@doi [The Astrophysical
  Journal] {10.1086/341225}, 571, L31

\bibitem[\protect\citeauthoryear{Yamazaki, Ioka  \& Nakamura}{Yamazaki
  et~al.}{2003}]{Yamazaki2003a}
Yamazaki R.,  Ioka K.,   Nakamura T.,  2003, \mn@doi [The Astrophysical
  Journal] {10.1086/376677}, 593, 941

\bibitem[\protect\citeauthoryear{Zhang}{Zhang}{2008}]{Zhang2008}
Zhang F.,  2008, \mn@doi [The Astrophysical Journal] {10.1086/590951}, 685,
  1052

\bibitem[\protect\citeauthoryear{Zhang \& Yan}{Zhang \& Yan}{2010}]{Zhang2011}
Zhang B.,  Yan H.,  2010, \mn@doi [The Astrophysical Journal]
  {10.1088/0004-637X/726/2/90}, 726, 90

\makeatother
\end{thebibliography}
}

\appendix

\section{Derivation of the pulse light curve and spectrum}

For the ease of the reader, and for notational clarity, in what follows, we reproduce some passages of the derivation of the formulas used for the pulse light curves and time dependent spectra. Such formulas are special cases of the more general formalism developed in \cite{1999ApJ...523..187W}.

\subsection{Light curve of the pulse from an expanding sphere}\label{sec:exp-sphere}
In general, the flux from a time-varying source can be defined as
\begin{equation}
 F(t) = \int_{S(t)}I(s)\cos\alpha\,ds/r^2
 \label{eq:flux-moving-source}
\end{equation} 
where $S(t)$ is the ``$t$-equal arrival time surface'' (EATS hereafter), \ie the locus of points of the source whose emitted photons reach the observer at $t$,  $r$ is the distance between the observer and the element $ds$ of the EATS, and $\alpha$ is the angle between the normal to the detector surface and the direction of the photon incoming from the $ds$ surface element. For most astrophysical applications $\cos\alpha\sim 1$, because the source is sufficiently far away to have a negligible angular size in the sky. Here the intensity $I(s)$ is allowed to vary both in time and in space ($s$ indicates the coordinates of a point on the EATS, which is a surface in spacetime), so this formulation is applicable to inhomogeneous sources as well.

In our situation, it is convenient to use spherical coordinates centred on the emitting sphere, so that (assuming cylindrical symmetry of the intensity) we have $ds = 2\pi\sin\theta\,d\theta\,R(t_e)^2$, where $t_e=t-r/c$ is the emission time. The distance $r$ of the point $(\theta,\phi,t_e)$ from the detector is $r\approx d-R(t_e)\cos\theta$ where $d$ is the distance of the sphere centre from the detector, thus
\begin{equation}
 F(t) = 2\pi\int_{S(t)} I(\theta,t_e)\sin\theta\,d\theta \dfrac{R(t_e)^2}{(d-R(t_e)\cos\theta)^2}
\end{equation} 
Since $d\gg R(t_e)$, the last term is well approximated by $R(t_e)^2/d^2$, thus we can write
\begin{equation}
 F(t) = \dfrac{2\pi}{d^2}\int_{S(t)} I(\theta,t_e)\,R(t_e)^2\,\sin\theta\,d\theta
 \label{eq:flux-partial}
\end{equation}

Let us now assume that the luminosity $L$ of the sphere is constant in the time interval $t_0<t<t_0+T$. This is different from assuming that the intensity is constant, in that it prevents the expansion of the surface area from causing a rise in the luminosity (this alternative assumption would be more appropriate in the description of an external shock). In terms of intensity, this assumption implies that $I\propto R^{-2}$, which we write as
\begin{equation}
 I(\theta,t_e) = I_0(\theta)\,\dfrac{R^2}{R^2(t_e)}
\end{equation} 
Inserting this definition into Eq.~\ref{eq:flux-partial} allows us to bring the radius outside the integral. In sections \S\ref{sec:eats-sphere} and \S\ref{sec:eats-exp-sphere}, we found that the EATS are the portions of the sphere comprised between $\cos\theta_{\rm on}$ and $\cos\theta_{\rm off}$, so that we have, for $t>0$

\begin{equation}
  F(t) = \dfrac{2\pi\,R^2}{d^2}\int_{\theta_{\rm off}(t)}^{\theta_{\rm on}(t)} I_0(\theta)\,\sin\theta\,d\theta
\end{equation}
Now, since the sphere is expanding, in the approximation of infinitesimal shell thickness, the intensity is related to the comoving one by $I_0(\theta)=\delta^4(\theta)\,I_0'$, where $\delta(\theta)=\left[\Gamma (1-\beta\cos\theta)\right]^{-1}$ is the Doppler factor, and $I_0'$ is the comoving intensity, assumed isotropic.
The flux is then
\begin{equation}
  F(t) = \dfrac{2\pi\,R^2}{d^2}\dfrac{I_0'}{\Gamma^4}\int_{\cos\theta_{\rm on}(t)}^{\cos\theta_{\rm off}(t)} \dfrac{d\cos\theta}{\left(1-\beta\cos\theta\right)^4}
  \label{eq:sphere-flux-integral}
\end{equation}
which yields, after substitution of the expressions for $\cos\theta_{\rm on}$ and $\cos\theta_{\rm off}$ derived above, the light curve of the pulse
\begin{equation}
 F(t) = F_{\rm max} \times \left\lbrace \begin{array}{lr}
                                    1 - \left(1+\dfrac{t}{\tau} \right)^{-3} & t\leq \toff\\
                                    \left(1+\dfrac{t-\toff}{\tau + t_{\rm off}} \right)^{-3} - \left(1+\dfrac{t}{\tau} \right)^{-3} & t>\toff
                                    \end{array}\right.\label{eq:pulse-sphere}
\end{equation}
where
\begin{equation}
 F_{\rm max}=\dfrac{2\pi\,R^2}{3 d^2}\dfrac{(1+\beta)^3\,\Gamma^2\,I_0'}{\beta}
\end{equation}
is the (saturation) peak flux if the pulse lasts $T\gg R/c$,
\begin{equation}
 \tau = \dfrac{R}{\beta c(1+\beta)\Gamma^2}
\end{equation} 
and 
\begin{equation}
 \toff  = \dfrac{T}{(1+\beta)\Gamma^2} = \dfrac{\Delta R}{\beta c(1+\beta)\Gamma^2} \equiv t_{\rm peak}
\end{equation} 
The fluence, \ie integral of the flux over time, from $t=0$ to $t\to \infty$ is
\begin{equation}
\mathcal{F} = F_{\rm max}\times\dfrac{3}{2}t_{\rm off} = \dfrac{\pi\,R^2}{d^2}\dfrac{(1+\beta)^2}{\beta}\,I_0'T 
\end{equation} 

It is worth noting that the light curve parameters are three, \ie $\tau$, $\toff$ and $F_{\rm max}$, while the underlying physical parameters are four, namely $R$, $T$, $\Gamma$ and $I_0'$. This degeneration leads to the impossibility to determine all the physical parameters by fitting the pulse shape to an observed light curve.

\subsection{Light curve of the pulse from an on-axis jet with $\tj\lesssim 1/\Gamma$}\label{sec:onaxis-jet-appendix}

To compute the pulse light curve of a jet of semiaperture $\theta_{jet}$, we can just take the pulse of the sphere and ``trim'' the unwanted part. If the jet is seen on-axis, this amounts to limit the integral of Eq.~\ref{eq:sphere-flux-integral} to angles $\theta < \theta_{jet}$. It is straightforward to work out at what time the EATS borders reach the jet border, \ie
\begin{equation}
 \theta_{\rm on}(t) = \theta_{jet} \implies t = R(1-\cos\theta_{jet})/c \equiv t_{\rm jet}
\end{equation} 
and similarly 
\begin{equation}
 \theta_{\rm off}(t) = \theta_{jet} \implies t = \toff + R_{\rm off}(1-\cos\theta_{jet})/c \equiv t_{\rm jet,off}
\end{equation}
where $R_{\rm off} \equiv R + \Delta R$.
It is then easy to see that the light curve becomes
\begin{equation}
 \dfrac{F(t)}{F_{\rm max}} =  1 - \left(1+\dfrac{\min(t,t_{\rm jet})}{\tau} \right)^{-3}
 \label{eq:pulse-thj-1-appendix}
\end{equation}
for $t\leq \toff$, then
\begin{equation}
   \dfrac{F(t)}{F_{\rm max}} = \left(1+\dfrac{t-\toff}{\tau_{\rm off}} \right)^{-3} - \left(1+\dfrac{\min(t,t_{\rm jet})}{\tau} \right)^{-3}
   \label{eq:pulse-thj-2-appendix}
\end{equation} 
for $\toff<t<t_{\rm jet,off}$, and zero for $t\geq t_{\rm jet,off}$.
This light curve is the same as that of the expanding sphere up to $t=t_{\rm jet}$. After that, if $t_{\rm jet}<\toff$ the flux saturates (the whole jet is visible) until $t=\toff$, then it drops and reaches zero at $t=t_{\rm jet,off}$. If $t_{\rm jet}\geq \toff$ no saturation is reached. The difference between the expanding sphere and the on-axis jet is relevant only if $t_{\rm jet}\lesssim \tau$, \ie if $\tj\lesssim 1/\Gamma$, as expected.

\subsection{Off-axis jet}\label{sec:offaxis-appendix}

\begin{figure}
 \begin{center}
 \includegraphics[width=\columnwidth]{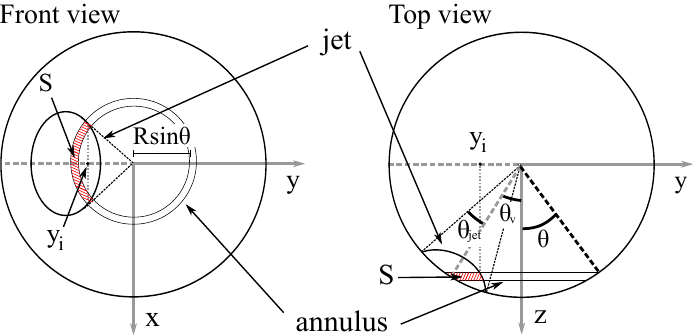} 
 \end{center}
 \caption{\label{fig:geom4}  The off-axis jet can be thought of as being part of an expanding sphere. The axes in the figures above are chosen so that the jet axis lies on the $z$-$x$ plane.  Jet surface elements in the $S$ shaded part all share the same Doppler factor $\delta$, and thus they all give the same contribution (per unit emitting area) to the flux. For this reason, the ratio of the flux from the annulus to the flux from $S$ is just equal to the ratio of the corresponding surface areas. \textbf{Left:} the jet is seen from the $z$-axis. The $y$-coordinate $y_i$ of the interceptions between the annulus and the jet border is shown. \textbf{Right:} the jet is seen from the $x$-axis. Angles $\tv$, $\tj$ and $\theta$ are reported.}
\end{figure}

If the jet is off-axis, it is still possible to compute an expression for the light curve. We propose here an approach to the computation, based on geometrical arguments.  Let us call $\tv$ the angle between the jet axis and the line of sight, $\tj$ the jet half-opening angle, and let us set the coordinate system so that the jet axis lies in the $z-x$ plane, as in Fig.~\ref{fig:geom4}. This is what one would obtain by rotating an on-axis jet counter-clockwise by an angle $\tv$ around the $x$ axis.
Let us now consider the ring-shaped part of the sphere surface (``annulus'' hereafter) comprised between $\theta$ and $\theta+d\theta$. If $\theta > |\tv-\tj|$, a portion $S$ of the annulus lies on the jet surface (shaded part in Fig.~\ref{fig:geom4}). Since the annulus width $d\theta$ is infinitesimal, the ratio of the area of $S$ to the total annulus area is equal to the ratio between the length $l$ of $S$ and the total annulus length $2\pi\,R\cos\theta$. Moreover, this is also equal to the ratio of the flux $dF_S$ from $S$ to the flux $dF_a$ from the whole annulus, namely
\begin{equation}
 \dfrac{dF_{S}(\theta)}{dF_{a}(\theta)} = \dfrac{l(\theta)}{2\pi\,R\cos\theta}
 \label{eq:ratios}
\end{equation} 

The flux due to the annulus is easily obtained by deriving the flux of the sphere, Eq.~\ref{eq:sphere-flux-integral}, with respect to $\theta$, which gives
\begin{equation}
 dF_{a}(\theta)=\dfrac{dF}{d\theta}d\theta=\dfrac{2\pi\,R^2}{d^2}\dfrac{I_0'}{\Gamma^4}\dfrac{\sin\theta\,d\theta}{\left(1-\beta\cos\theta\right)^4}
 \label{eq:fannulus}
\end{equation}

To compute the length $l(\theta)$, we must first find the interceptions between the annulus and the jet border. Both are circles on the sphere surface, \ie they lie on the surface $x^2+y^2+z^2=R^2$. The annulus is the circle given by the interception between the plane $z=R\cos\theta$ and the sphere; in a coordinate system $K'$ where the $z'$ axis coincides with the jet axis, the jet border is the circle given by the interception between the plane $z'=R\cos\tj$ and the sphere. Applying a rotation of an angle $\tv$ around the $x$ axis, this plane becomes $z\cos\tv - y\sin\tv = R\cos\tj$. The interceptions between the two circles are then found by solving the linear system
\begin{equation}
 \left\lbrace\begin{array}{l}
              x^2+y^2+z^2=R^2\\
              z=R\cos\theta\\
              z\cos\tv - y\sin\tv = R\cos\tj\\
             \end{array}\right.
\end{equation} 
The $y$ coordinate of the interceptions (see Fig.~\ref{fig:geom4}) is found to be
\begin{equation}
 y_i = \dfrac{\cos\theta\cos\tv-\cos\tj}{\sin\tv}R
\end{equation}
Consider now the annulus as a circle whose center lies on the $z=R\cos\theta$ plane. Its radius is $R\sin\theta$, and the angle $\alpha$ that subtends S is $\alpha = 2\arccos\left(-y_i/R\sin\theta\right)$.  The length $l(\theta)$ is then
\begin{equation}
 l(\theta) = 2 R \sin\theta \arccos\left(\dfrac{\cos\tj - \cos\theta\cos\tv}{\sin\theta\sin\tv}\right)
 \label{eq:l}
\end{equation}
Substituting Eqs.~\ref{eq:fannulus} and \ref{eq:l} into Eq.~\ref{eq:ratios}, we conclude that
\begin{equation}
 dF_S(\theta) = \dfrac{dF}{d\theta}d\theta\times\dfrac{1}{\pi}\arccos\left(\dfrac{\cos\tj - \cos\theta\cos\tv}{\sin\theta\sin\tv}\right)
\end{equation} 
This is valid as long as the interceptions between the annulus and S exist, \ie for $|\tv-\tj|<\theta<\tv+\tj$. Let us work out the remaining cases:
\begin{itemize}
 \item if $\tv<\tj$, \ie if line of sight is inside the jet border, then for $\theta<\tj-\tv$ also the annulus is inside the jet border, thus $dF_S(\theta) = dF_a(\theta)$;
 \item if $\tv>\tj$, \ie if line of sight is outside the jet border, then for $\theta<\tv-\tj$ the annulus is too small to intercept the jet border, thus $dF_S(\theta)=0$;
 in either case, if $\theta>\tv+\tj$ the annulus is too large to intercept the jet border, thus again $dF_S(\theta)=0$.
\end{itemize}

Summing up, we can define the function $a(\theta,\tv,\tj)$ by
\begin{equation}
 a = \left\lbrace\begin{array}{lr}
                          H(\tj-\tv) & \theta\leq|\tv-\tj|\\
                          0 & \theta\geq\tv+\tj\\
                          \dfrac{1}{\pi}\arccos\left(\dfrac{\cos\tj - \cos\theta\cos\tv}{\sin\theta\sin\tv}\right) & \rm otherwise\\
                         \end{array}\right.
\end{equation} 
where $H(x)$ is the Heaviside function, \ie
\begin{equation}
 H(x) = \left\lbrace\begin{array}{lr}
         0 & x<0\\
         1 & x\geq 0\\
        \end{array}\right.
\end{equation}
and write
\begin{equation}
 dF_S(\theta,\tv,\tj) = a(\theta,\tv,\tj)\dfrac{dF}{d\theta}d\theta
 \label{eq:dF_S}
\end{equation}
The light curve of the pulse from the off-axis jet is then obtained by integration of this expression between $\theta_{\rm on}(t)$ and $\theta_{\rm off}(t)$, namely
\begin{equation}
 F(t,\tv,\tj) = \dfrac{2\pi\,R^2}{d^2}\dfrac{I_0'}{\Gamma^4}\int_{\theta_{\rm off}(t)}^{\theta_{\rm on}(t)}a(\theta,\tv,\tj)\dfrac{\sin\theta\,d\theta}{\left(1-\beta\cos\theta\right)^4}
 \label{eq:flux-offaxis}
\end{equation} 
Note that here $t=0$ is the arrival time of the first photon from the sphere, thus if $\tv>\tj$ the actual light curve of the off-axis jet starts a little later. The actual start time of the off-axis light curve is
\begin{equation}
 t_{\rm start}(\tv,\tj) = \dfrac{R}{c}\left(1-\cos(\tv-\tj)\right)
\end{equation} 
Equation~\ref{eq:flux-offaxis} can be easily integrated with a simple numerical procedure. Some example light curves computed using a RK4 integration scheme are given in Fig.~\ref{fig:homogeneous-off-axis-lcs}. 

\subsection{Spectra}
All the above arguments can be also applied to the derivation of the observed spectrum. All we need to do is to compute the flux density
\begin{equation}
 \dfrac{dF}{d\nu}(\nu,t) \equiv F_{\nu}(\nu,t)=  \int_{S(t)} \dfrac{dI}{d\nu}(\nu,t)\cos\alpha\,ds/r^2
\end{equation}
over the same EATS as before, using the transformation
\begin{equation}
 \dfrac{dI}{d\nu}(\nu) = \delta^3\,\dfrac{dI'}{d\nu'}(\nu/\delta)
\end{equation}
to express the intensity density in terms of the comoving one.
It is convenient to write $dI'/d\nu'$ as follows
\begin{equation}
 \dfrac{dI'}{d\nu'}(\nu') = \dfrac{I_0'}{\nu_0'}f(\nu'/\nu_0')
\end{equation}
where $I_0'$ is the total intensity, $\nu_0'$ is some frequency, and $f(x)$ is a function which describes the comoving spectral shape, and whose integral is normalized to unity. As an example, we can set a power law spectral shape
\begin{equation}
 f(\nu'/\nu_0') = \left(1-\alpha\right)\left( \dfrac{\nu'}{\nu_0'} \right)^{-\alpha}
\end{equation}
for $\nu'>\nu_0'$ and zero otherwise, with $\alpha>1$. Since the integral of $f(x)$ is normalized to unity, we have
\begin{equation}
 \int_{0}^{\infty}\dfrac{dI'}{d\nu'}(\nu')d\nu' = \dfrac{I_0'}{\nu_0'}\int_{0}^{\infty}f(\nu'/\nu_0')d\nu'=I_0'
\end{equation}
The equation for the observed spectrum of a off-axis jet is then
\begin{equation}
 \dfrac{dF}{d\nu}(\nu,t) = \dfrac{2\pi\,R^2}{d^2}\dfrac{I_0'}{\nu_0'\Gamma^3}\int_{\theta_{\rm off}(t)}^{\theta_{\rm on}(t)}a(\theta,\tv,\tj)\dfrac{f(\nu/\delta\,\nu_0')\sin\theta\,d\theta}{\left(1-\beta\cos\theta\right)^3}
 \label{eq:off-axis-spectrum}
\end{equation}

For the simplest case of an on-axis jet, with power law comoving spectral shape, the integral is analytic and it gives
\begin{equation}
\begin{array}{l}
\dfrac{dF}{d\nu}(\nu,t) = \dfrac{2\pi R^2}{d^2}\dfrac{I_0'\left( 1+\beta \right)^{2+\alpha}\Gamma^{1+\alpha}}{\beta\,\nu_0'^{1-\alpha}}\left( 1-\alpha \right)\nu^{-\alpha}\times \\
\times \left\lbrace \left( 1 + \dfrac{t-\toff}{\tau + \toff} \right)^{-2-\alpha} - \left(1+\dfrac{t}{\tau}\right)^{-2-\alpha} \right\rbrace\\
\end{array}
\end{equation}
which reproduces the well-known $2+\alpha$ decay slope due to high latitude emission \citep{Kumar2000,Dermer2004}.
For more general spectral shapes, a numerical approach is necessary to compute the integral in Eq.~\ref{eq:off-axis-spectrum}. For this paper, in most cases a IV order Runge-Kutta method has been used to compute separately the specific flux light curve at a number of frequencies. The values of the specific fluxes at a certain time then constitute the spectrum at that time.

 \label{lastpage}
 \end{document}